\documentclass[%
 reprint,
 superscriptaddress,
 amsmath,amssymb,
 aps,
]{revtex4-2}

\usepackage{graphicx}
\usepackage{dcolumn}
\usepackage{bm}
\usepackage{color}
\usepackage{float}

\begin{document}

\preprint{APS/123-QED}

\title{
    Chemical Equilibration and Thermalization of Quark-Gluon Plasma\\
    in a Parton Cascade Model with 2-to-3 Quark Interactions
}

\author{Cendikia Abdi}
    \affiliation{Department of Physics, Hiroshima University, Higashihiroshima 739-8526, Japan.}
\author{Chiho Nonaka}
    \affiliation{Department of Physics, Hiroshima University, Higashihiroshima 739-8526, Japan.}
    \affiliation{Department of Physics, Nagoya University, Nagoya 464-8602, Japan.}
    \affiliation{Kobayashi Maskawa Institute, Nagoya University, Nagoya 464-8602, Japan.}

\date{\today}

\begin{abstract}
We investigate the thermalization, chemical equilibration, and hydrodynamization behavior of the far-from-equilibrium, gluon-dominated quark gluon plasma (QGP) produced in Au+Au collisions at $\sqrt{s_{\text{NN}}} = 200$ GeV using the hadronic transport model SMASH extended to simulate partonic interactions. The initial conditions are prepared using the mini-jet model with nuclear parton distribution functions. We first validate the model in a box simulation with the periodic boundary condition to establish indicators for thermalization, chemical equilibration, and hydrodynamization by analyzing energy spectrum and momentum anisotropy. We observe that the additional inelastic channels accelerate thermalization and chemical equilibration compared to the gluon-only scheme. Applying the same framework to the expanding medium, we find that the energy spectrum converges toward the Boltzmann distribution at $t \sim 0.2$ fm while momentum isotropization is achieved at $t \sim 2$ fm, but chemical equilibration is not clearly established even after 5 fm. The Knudsen number rises above unity after $\sim 4$ fm, indicating a breakdown of the hydrodynamic regime at later times consistent with other kinetic theory approaches.
\end{abstract}

\maketitle

\section{\label{sec:level1} Introduction}
The creation of Quark-Gluon Plasma (QGP) in relativistic heavy-ion collisions provides a controllable environment for probing the fundamental properties of matter under extreme conditions. The most remarkable finding from experiments at the Relativistic Heavy Ion Collider (RHIC) and the Large Hadron Collider (LHC) is that QGP behaves as a near-perfect fluid \cite{Gyulassy:2004, Cifarelli:2024kgh}. Meanwhile, the transition from the initial, highly non-equilibrium gluon matter to a regime governed by collective hydrodynamic expansion, as suggested by the hydrodynamic model happens in a very short time around 0.2-1.0 fm after the collision. This rapid hydrodynamization and equilibration process remains as a significant gap in our theoretical understanding of heavy-ion collisions \cite{Keegan2016, PhysRevD.99.054018, PhysRevD.46.4986, PhysRevLett.122.122302, BAIER200151}. Therefore, understanding the dual processes of hydrodynamization and equilibration is paramount.

In general, it has been shown that hydrodynamic model is essential to simulate the time evolution of the QGP to reproduce the collective behavior observables from high-energy heavy-ion collision experiments \cite{BACK200528, ADCOX2005184, ARSENE20051, Petersen:2010cw, Qiu:2011hf, PhysRevLett.116.132302, Teaney:2009qa, Zhao:2017rgg, Pandey2024-ts, Giacalone:2018cuy, BOZEK2015135, Goldschmidt:2015kpa, PhysRevC.98.054906, PhysRevC.107.014901, PhysRevC.107.034912}. This means that the properties of the quark and gluon particles in each fluid cell are represented as hydrodynamic and thermodynamic properties. However, such an assumption is only valid when the particles inside that certain volume are in an equilibrium state, thus rendering such an approach invalid for early time QGP due to its far-from-equilibrium nature.

Setting aside the question of how equilibrium is defined, the system begins in a highly anisotropic and energetic non-equilibrium state \cite{Moreland:2014oya, Schenke:2012wb, Schenke:2016ksl}, it then eventually hydrodynamizes and equilibrates \cite{Strickland:2017kux, Kurkela:2018wud, Boguslavski:2023jvg}. As such, hydrodynamic simulation does not account for the pre-equilibrium state and is initialized already in the equilibrium state a certain time after the collision. Another issue is that there is still no clear definition as to when the equilibrium is reached to the point where hydrodynamic calculation is applicable.

The nuclear matter created after the collision starts as a highly energetic non-equilibrium matter and because the properties of its particles cannot be represented in thermodynamic parameters, thus equilibrium statistical physics cannot be applied in this phase. Moreover, it has been shown that the description of the pre-equilibrium QGP does play an important role in determining resulting observables both for heavy energetic partons \cite{Das:2017dsh} as well as the partons in the soft region \cite{Pablos:2022piv}. Thus it was clear that it is crucial to describe this phase carefully. 

The common approach to determine the initial state in a hydrodynamic model often uses free streaming \cite{CHATTOPADHYAY2020135158} or includes assumptions to simplify the pre-equilibrium state of QGP. Despite the lack of a realistic model to describe the pre-equilibrium phase, hydrodynamic models have been shown to be able to reproduce many observable data well \cite{Nijs:2023yab}. However, this free streaming approach does not align with the strongly interacting nature of QCD.

Recently, there have been various models to incorporate the strongly coupled behavior of QCD in the pre-equilibrium phase from simple bottom-up equilibration in a 1D expanding medium from perturbative QCD based on the Yang-Mills transport equation \cite{PhysRevD.103.056010}, to the effective kinetic theory approach based on the Boltzmann transport equation \cite{Tornkvist:2023ylh}. There is also a phenomenological model to simulate equilibrium and pre-equilibrium states at the same time and introduces the interplay between the pre-equilibrium partons and equilibrium QGP \cite{Kanakubo:2021qcw}. On the other hand, there were also studies from the hydrodynamic approach, such as the anisotropic hydrodynamic model \cite{Chen:2024grb}, effective non-equilibrium hydrodynamic model and its connection to holography \cite{Kaminski:2025ika}, or matching the parametrized state at the end of the pre-equilibrium stage and the initial state of relativistic viscous hydrodynamic \cite{Kurkela:2018wud}. In this study, we opt to employ kinetic theory with the Boltzmann transport equation in a full 3D simulation because it does not suffer from the same limitation as hydrodynamic calculation.

One of the main goals of this study is to find reliable indicators of these processes. Equilibration is divided into two processes, namely kinetic equilibration and chemical equilibration. Kinetic equilibration refers to the process of the QGP to approach thermal equilibrium often defined based on how close the energy spectrum to the Maxwell-Boltzmann distribution \cite{Xu:2004mz, BarreraCabodevila:2023apf}. On the other hand, the traditional definition of chemical equilibrium is the balance in the particle production and annihilation for each species, thus resulting in the saturation of particle numbers. Nevertheless, past studies had also investigated chemical equilibration based on the partial energy density \cite{Du:2020jhm} or particle number ratio between gluons and quarks \cite{PhysRevLett.122.142301}. In contrast, hydrodynamization is a process where the time evolution behavior saturates toward a system described by hydrodynamic equation of motion regardless of the initial condition of the system. This saturation behavior is often parametrized by hydrodynamic attractors such as the degree of anisotropicity \cite{Kurkela:2015qoa} or the energy density \cite{Kurkela:2018xxd}.

Regardless of the baseline used to determine equilibration and hydrodynamization, it has been a common problem among kinetic theory approaches where hydrodynamization takes 2 - 5 fm while on the other hand, the initialization time often used in hydrodynamic models is around 0.2-1.0 fm. This study is motivated by the original study with BAMPS and 2-to-3 interaction exclusive to gluon\cite{Xu:2004mz}, which still results in longer equilibration and hydrodynamization time than expected. Our work focuses on expanding the 2-to-3 interaction to include quarks as well considering that there is a strong quark production at the early time and there is a switch between a gluon-dominated and a quark-dominated system to achieve chemical equilibrium. For the purpose of comparing the effect of the extra interaction channels, we focus on the central collision of Au-Au at RHIC energy level.

In this paper, we first describe the details of the numerical simulation SMASH employed throughout the calculation in Sec.~\ref{sec:smash_introduction} including the modifications introduced to simulate partonic system. In Sec.~\ref{sec:box_test} we simulate a simple system in a box with the periodic boundary condition already initialized in equilibrium state, we then ensure that the equilibrium state is stable and that the screening mass yields the expected value for a system in equilibrium. After verifying that the equilibrium system behaves as expected, in Sec.~\ref{sec:box_equilibration} we initialize the system in a far-from-equilibrium state in a box with the periodic boundary condition. We then observe how the system approaches equilibration and hydrodynamization to establish suitable indicators for both processes. Once we set up the basis to identify equilibration and hydrodynamization, we shall apply the same criteria in the heavy-ion collision case in Sec.~\ref{sec:hic_equilibration} and determine the hydrodynamization, thermalization, and chemical equilibration timescale. Finally, the summary of this paper and a brief comparison to the results from past studies are given in Sec.~\ref{sec:conclusion}.

\section{Transport Model SMASH}\label{sec:smash_introduction}
We modified the hadronic transport model SMASH~\cite{SMASH:2016zqf} to accommodate partonic interaction by adding partonic particles and cross-sections. The focus of this model is to solve the Boltzmann transport equation
\begin{equation}
	\frac{\partial}{\partial t} f(t, \vec{r}, \vec{p}) + \frac{\vec{p}}{E}\nabla_{\vec{r}} f(t, \vec{r}, \vec{p}) - \nabla_{\vec{r}} U(\vec{r}) \nabla_{\vec{p}} f(t, \vec{r}, \vec{p}) = \left( \frac{\partial f}{\partial t} \right)_{\text{coll}}.
\end{equation}
The Boltzmann transport equation describes the time evolution of the distribution function ($f$) due to the diffusion, effect from the external field ($U$), and inter-particle collision. Depending on how we define the collision kernel on the right hand side of the equation, we often end up with a non-linear partial differential equation with no known exact solution. The common approach is to make assumption or approximation to solve the equation. In this case, the transport model gives an approximate solution to the Boltzmann transport equation through the Monte-Carlo method. The Boltzmann transport equation is the basis of the kinetic theory calculation which has been used on several occasions in previous models \cite{PhysRevC.72.064901, ZHANG1998193, COLEMANSMITH2013759c}, a full analysis of hadronic gas based on Monte-Carlo approximate solution to Boltzmann transport equation shows a good agreement with theoretical prediction of hadronic gas thermodynamic properties \cite{Torbjorn_Sjostrand_2006, M_Bleicher_1999}.

Although it is already argued that a strong external field exists in the early stage of the QGP which affects physical observables \cite{PhysRevLett.134.172301, PhysRevC.109.034917, Yan:2021zjc}, for the sake of simplicity, the simulation throughout this paper is done with no external field ($U = 0$). Moreover, due to SMASH is originally a hadron transport model, the algorithm has to be reconfigured to simulate parton-parton interactions. In contrast to the hadronic case, there are no experimental data to ascertain parton-parton cross section. Hence, we opt to determine the partonic interaction cross section directly from perturbative QCD (pQCD) calculation \cite{Abir:2010kc, PhysRevD.25.746, PhysRevD.88.014018, Fochler_2013} and fix the coupling constant to $\alpha_s = 0.3$ throughout this study.

\subsection{Stochastic Collision Criterion}
As we include 2-to-3 interaction, the backward direction process is necessary to satisfy the detailed balance principle. The geometric collision criterion decides inter-particle collision based on the distance between two nearby particles; as such, the same criterion cannot be used to identify many-body interactions. In order to include 3-to-2 interaction, we adopt the stochastic collision criterion described in Refs.~\cite{PhysRevC.104.034908, PhysRevC.97.044907, BUSS20121, LANG1993391} as,
\begin{equation}
    \begin{aligned}
        P_{2 \rightarrow \text{n}} =& \frac{\Delta t}{\Delta^{3} x} \nu_{\text{rel}} \sigma_{2 \rightarrow \text{n}}(s),\\
        P_{3 \rightarrow 2} =& \left( \frac{g'_{1}g'_{2}}{g_{1}g_{2}g_{3}} \right) \frac{S!}{S'!} \frac{\Delta t}{\Delta^{3} x} \frac{E'_{1}E'_{2}}{2E_{1}E_{2}E_{3}} \frac{\Phi_{2}(s)}{\Phi_{3}(s)} \\
        & \times \nu_{\text{rel}} \sigma_{2 \rightarrow 3}(s),
    \end{aligned}
\end{equation}
respectively for 2-to-n and 3-to-2 processes. Here $g$'s are the degeneracy factors for each outgoing or ingoing particle, $S$ and $S'$ are the symmetric factors for each process, $\Phi_{\text{n}}$ is the n-body phase space integration, and $s$ is the collision energy.

\subsection{2-to-2 Cross Section}
The 2-to-2 cross sections are calculated using perturbative QCD up to tree level including all three Mandelstam variables $s$, $t$, and $u$ channels, which are explicitly given as,
\begin{equation}\label{eq:full_cross_2_to_2}
    \begin{aligned}
        \frac{d\sigma_{gg \rightarrow gg}}{dt} =& \frac{9 \pi \alpha^{2}_{s}}{2s^{2}} \left[3 - \frac{tu}{s^{2}} - \frac{su}{t^{2}} - \frac{st}{u^{2}} \right],\\
        \frac{d\sigma_{gq \rightarrow gq}}{dt} =& \frac{4 \pi \alpha^{2}_{s}}{9s^{2}} \left[ - \frac{u}{s} - \frac{s}{u} + \frac{9(s^{2} + t^{2})}{4t^{2}} \right],\\
        \frac{d\sigma_{qq \rightarrow qq}}{dt} =& \frac{4 \pi \alpha^{2}_{s}}{9s^{2}} \left[ \frac{u^{2} + s^{2}}{t^{2}} + \frac{t^{2} + s^{2}}{u^{2}} - \frac{2s^{2}}{3ut} \right],\\
        \frac{d\sigma_{gg \rightarrow q \bar{q}}}{dt} =& \frac{\pi \alpha^{2}_{s}}{6s^{2}} \left[ \frac{u}{t} + \frac{t}{u} - \frac{9(t^{2} + u^{2})}{4s^{2}} \right],\\
        \frac{d\sigma_{q \bar{q} \rightarrow gg}}{dt} =& \frac{8^{2}}{3^{2}} \frac{d\sigma_{gg \rightarrow q \bar{q}}}{dt}.
    \end{aligned}
\end{equation}
The full cross section for a given collision energy $s$ is then calculated by integrating the Mandelstam variable $t$ over all the possible values of energy exchange
\begin{equation}
    \sigma_{2 \rightarrow 2} = \int_{0}^{P_{\text{cms}}} \frac{d \sigma_{2 \rightarrow 2}}{dt} dt,
\end{equation}
where  
    $P_{\text{cms}} = ({s^{2} + m_{1}^{4} + m_{2}^{4} - 2 (m_{1}^{2}s + m_{2}^{2}s + m_{1}^{2}m_{2}^{2})})/s$
is the kinematic upper limit of the exchanged energy between the incoming particles in the center of mass frame. In case of the massless limit of u, d, and s quarks, we have $P_{\text{cms}} \rightarrow s$. The lower limit of the integration starts from 0, which causes infrared divergence due to the $1/t$ terms in the differential cross section. In order to avoid the divergence, we applied dynamic screening mass to be discussed in subsection \ref{subsec:screening_mass}.

\subsection{2-to-3 Cross Section}
The 2-to-3 cross section is based on improved Gunion-Bertsch approximation \cite{PhysRevD.25.746, PhysRevD.88.014018} which was extended for heavy quark case \cite{Uphoff_2015}. In case of $qg \rightarrow qgg$, the improved Gunion-Bertsch approximation gives interaction amplitude as,
\begin{equation}
\begin{aligned}
&|M_{qg \rightarrow qgg}|^{2} = 12g^{2}|M_{qg \rightarrow qg}|^{2} (1 - \bar{x})^{2} \\
    \times & \left[ \frac{\vec{k}_{\perp}}{k_{\perp}^{2} + x^{2}M^{2} + m_{D}^{2}} + \frac{\vec{q_{\perp}} - \vec{k}}{(\vec{q_{\perp}} - \vec{k_{\perp}})^{2} + x^{2}M^{2} + m_{D}^{2}} \right]^{2},
\end{aligned}
\end{equation}
with $k_{\perp}$ is the transverse momentum of the radiated gluon and $q_{\perp}$ is the transverse momentum of the exchanged gluon. The Gunion-Bertsch approximation effectively breaks down $|M_{2 \rightarrow 3}|$ amplitude as $|M_{2 \rightarrow 2}|$ amplitude times $|M_{1 \rightarrow 2}|$ gluon splitting amplitude under the soft gluon exchange and soft gluon radiation expressed as,
\begin{equation}
\begin{aligned}
    &k_{\perp} \ll \sqrt{s},  \\
    &q_{\perp} \ll \sqrt{s}.
\end{aligned}
\end{equation}
As discussed in Ref.~\cite{Fochler_2013}, the factor
\begin{equation}
\bar{x} = \frac{k_{\perp}}{\sqrt{s}} e^{|y|}    
\end{equation}
is crucial to correctly replicate the result from the exact calculation of $2 \rightarrow 3$ interaction amplitude as the original Gunion-Bertsch approximation overestimates the differential cross section in the backward momentum rapidity phase space of the radiated gluon. This is also one of the key improvements of the current model over previous kinetic theory calculations for the equilibration process in high-energy collision systems \cite{Xu:2004mz}. The same formula is also applicable for light quark case since for $M \rightarrow 0$, the interaction amplitude returns to the original improved Gunion-Bertsch approximation calculated for massless case \cite{Fochler_2013}.

In order to maintain consistency, we follow the assumptions imposed in the process of deriving the Gunion-Bertsch approximation in which the amplitude for 2-to-2 is from the small-angle approximation instead of the full amplitude. Thus $|M_{2 \rightarrow 2}|$ here is given as,
\begin{equation}\label{eq:cross_sec_2_to_2}
    \begin{aligned}
        |M_{gg \rightarrow gg}|^{2}_{\text{sa}} =& \frac{2}{9} g^{4} \frac{4s^{2}}{t^{2}},\\
        |M_{qg \rightarrow qg}|^{2}_{\text{sa}} =& \frac{4}{9} |M_{gg \rightarrow gg}|^{2}_{\text{sa}},\\
        |M_{qq \rightarrow qq}|^{2}_{\text{sa}} =& \frac{16}{81} |M_{gg \rightarrow gg}|^{2}_{\text{sa}}.
    \end{aligned}
\end{equation}
We focus on these three amplitudes as we include $gg \rightarrow ggg$, $qg \rightarrow qgg$, and $qq \rightarrow qqg$ channels.

Thus, the total cross section is then calculated as
\begin{equation}
    \begin{aligned}
        \sigma_{2 \rightarrow 3} =& \frac{1}{256 \pi^{4}} \frac{1}{S!} \frac{1}{s} \int_{0}^{\frac{s}{4}} dq_{\perp}^{2} \int_{0}^{\frac{s}{4}} dk_{\perp}^{2} \int_{y_{\text{min}}}^{y_{\text{max}}} dy \int_{0}^{\pi} d \phi \\
        & \times |M_{2 \rightarrow 3}|^{2} \sum \left( \left.\frac{\partial F}{\partial y_{3}}\right|_{F = 0} \right)^{-1},
    \end{aligned}
\end{equation}
where
\begin{equation}
    \begin{aligned}
        F =& s - 2 \sqrt{s} [q_{\perp}\cosh{(y_{3})} + k_{\perp} \cosh{(y)}] + 2q_{\perp}k_{\perp} \cos{(\phi)}\\
        & + 2q_{\perp}k_{\perp}[\cosh{(y_{3})} \cosh{(y)} - \sinh{(y_{3})} \sinh{(y)}]
    \end{aligned}
\end{equation}
and 
\begin{equation}
    y_{\text{max/min}} = \pm \text{arccosh} \left( \frac{\sqrt{s}}{2k_{\perp}} \right),
\end{equation}
with $\nu$ is the symmetric factor of each process, $q_{\perp}$ and $k_{\perp}$ are the perpendicular momenta of the exchanged and radiated gluon relative to the collision axis in the center of mass frame, $y_{3}$ is the rapidity of the radiated gluon, and $\phi$ is the angle between $q_{\perp}$ and $k_{\perp}$.

\subsection{Screening Mass}\label{subsec:screening_mass}
As shown in Eq.~(\ref{eq:cross_sec_2_to_2}), pQCD-cross sections suffer from infrared divergence when the Mandelstam variable $t$ is zero. Hence, in order to avoid this problem, we employ dynamic screening mass in one-loop approximation in SU(3) up to $\alpha_{s}$ order \cite{PhysRevC.54.2588},
\begin{equation}\label{eq:screening_mass}
\begin{aligned}
    &m_{D}^{2} = 16 \pi \alpha_{s} \int \frac{d^{3}p}{(2 \pi)^{3}} \frac{1}{|p|} (N_{c}f_{g} + n_{f}f_{q}), \\
    &m_{q}^{2} = 4 \pi \alpha_{s} \frac{N_{c}^{2} - 1}{2N_{c}}  \int \frac{d^{3}p}{(2 \pi)^{3}} \frac{1}{|p|} (f_{g} + f_{q}),
\end{aligned}
\end{equation}
where $N_{c}$ is the number of color and $n_{f}$ is the number of quark flavor in the system.

Since parton cascade model describes the evolution of the system as a discrete system, we use the discrete form of the dynamic screening mass given as
\begin{equation}
\begin{aligned}
    & m_{g}^{2}=&&16 \alpha_{s} \pi \frac{1}{V} \\
    & && \times \left[ \frac{N_{c}}{2(N_{c}^{2}-1)}  \sum\limits_{i=1}^{N_{\text{gluon}}} \frac{1}{|p_{i}|} + \frac{1}{g_{q} N_{c} n_{f}} \sum\limits_{i=1}^{N_{\text{quark}}} \frac{1}{|p_{i}|}  \right], \\
    & m_{q}^{2}=&&4 \alpha_{s} \pi \frac{N_{c}^{2} - 1}{2N_{c}} \frac{1}{V} \\
    & && \times \left[ \frac{1}{2(N_{c}^{2}-1)}  \sum\limits_{i=1}^{N_{\text{gluon}}} \frac{1}{|p_{i}|} + \frac{1}{g_{q} N_{c} n_{f}} \sum\limits_{i=1}^{N_{\text{quark}}} \frac{1}{|p_{i}|}  \right],
\end{aligned}
\end{equation}
where $g_{q} = 4$ is the degree of freedom of quark from spin and quark-antiquark pair. Although the original derivation assumes an equilibrium state, we assume the same definition can be applied to a non-equilibrium system.

\subsection{Cross Section Table}
One of the main challenges of quantum molecular dynamics simulation is the high computational cost. This is due to the nature of the simulation technique which is designed to fully track the momentum and coordinate of each individual particle. Therefore, when we consider 2-to-2, 2-to-3, and 3-to-2 inter-particle interactions, the number of interaction probabilities is roughly proportional to $N(N-1)(N-2)$ with $N$ being the number of particles.

Thus, in order to shorten the simulation time, we employ the look-up table method for both 2-to-2 and 2-to-3 cross sections. The data points of the table are pre-calculated in the screening mass and collision energy range of
\begin{equation}
\begin{aligned}
    M_{\text{screening}}(\text{GeV}) &\in [0.1, 1], \\
    \sqrt{s}(\text{GeV}) &\in [0.075, 10].
\end{aligned}
\end{equation}
Bilinear interpolation/extrapolation is then used to calculate the cross-section based on these data points.

\begin{figure}[h]
  \includegraphics[width=\linewidth]{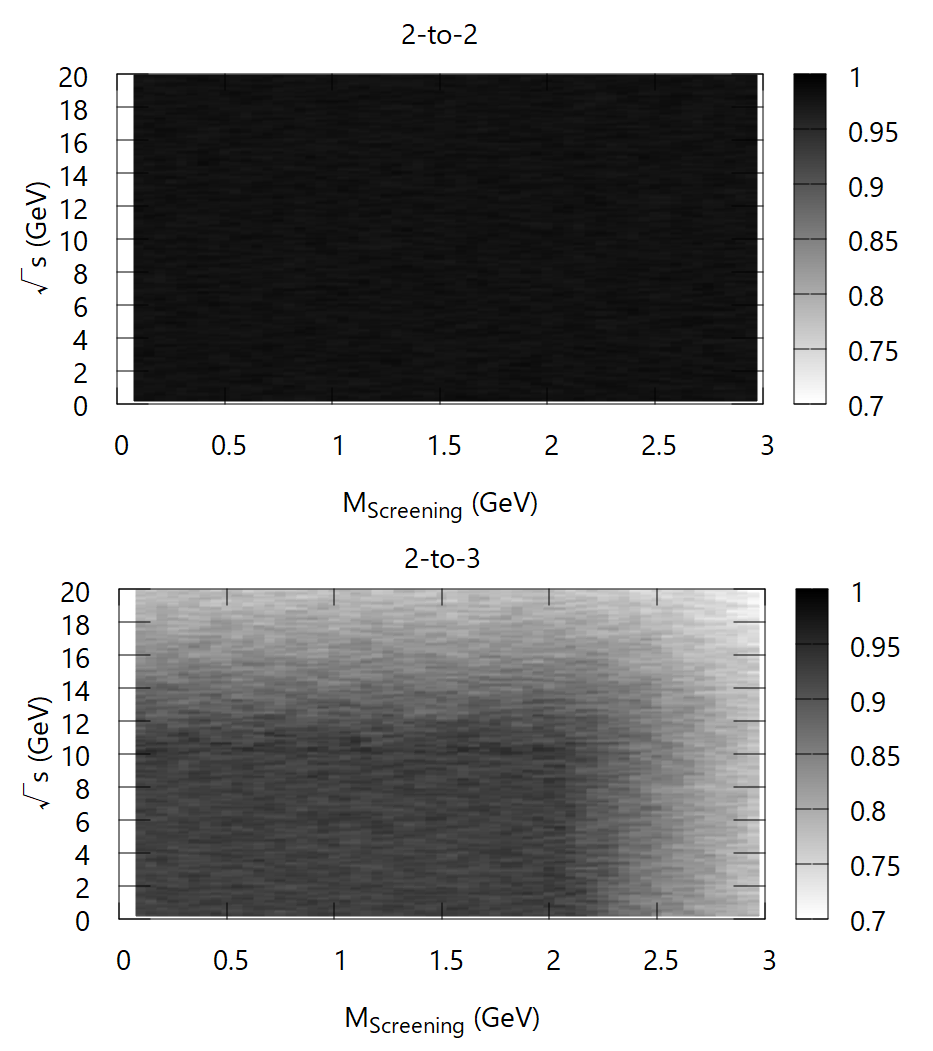}
  \caption{Cross section ratio between table interpolation and exact calculation for $gg \rightarrow gg$ (top) and $gg \rightarrow ggg$ (bottom) respectively. the rest of $2 \rightarrow 2$ and $2 \rightarrow 3$ channels have closely identical accuracy profile. Do note that the contour range is clamped between 0.7 and 1.0.}
  \label{fig:cross-section-acc}
\end{figure}

Figure \ref{fig:cross-section-acc} shows the ratio between the exact calculation of $gg \rightarrow gg$ and $gg \rightarrow ggg$ cross-section compared to the result calculated using a lookup table and bilinear interpolation/extrapolation up to $M_{\text{screening}} = 2$ GeV and collision energy $ \sqrt{s} = 20$ GeV. Similar behavior is also exhibited by the cross section for the other 2-to-2 and 2-to-3 channels. In the case of the 2-to-2 cross-section, we see accuracy is well above 97 \% across all the screening mass and collision energy range even outside the original data point range of the table. This is due to the 2-to-2 saturates quickly as we go to the higher collision energy region. On the other hand, in the case of 2-to-3 cross-section, we see a clear decrease of accuracy outside the table range as low as 70 \% around $M_{\text{screening}} = 2$ GeV or $ \sqrt{s} = 20$ GeV regions due to the fact that 2-to-3 cross section is proportional to $s^{2}$ at large s region, thus bilinear extrapolation yields a lower value compared to the quadratic increase of cross section with respect of collision energy. Nevertheless, the algorithm keeps accuracy well above 90 \% inside the table range. Due to this behavior, it is important to confirm that the inter-particle collision energy and screening mass are inside the region where the cross-section maintains a reasonable level of accuracy across the time evolution of the system.

\section{Box Simulation in Equilibrium State}\label{sec:box_test}
\subsection{Screening Mass}
To verify that we have applied the dynamic screening mass correctly, we prepared a box with a fixed 5 fm box length and periodic boundary conditions. For the purpose of simulating a system already in an equilibrium state, the initial momentum and the number density of the quarks and gluons are determined by the Boltzmann distribution function. While the analytic prediction is calculated by substituting the Boltzmann distribution function $f(E) = e^{- \frac{E}{T}}$ both for the quark and gluon distribution functions directly to Eq.~(\ref{eq:screening_mass}). 

\begin{figure}[h]
  \includegraphics[width=\linewidth]{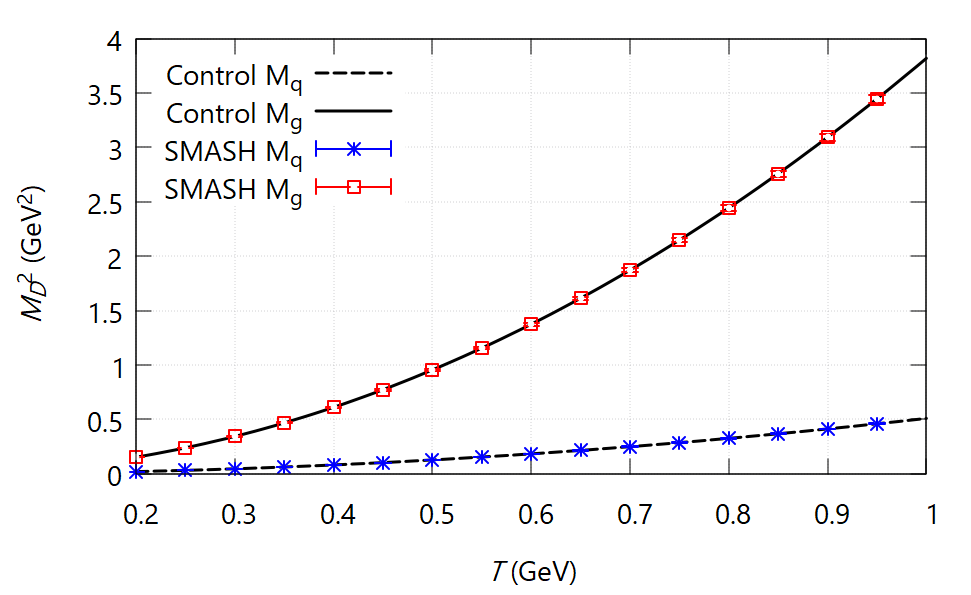}
  \caption{Screening mass comparison extracted from energy spectrum. Black lines stand for analytic calculation assuming Boltzmann distribution with the dashed line for quark and solid line for gluon. SMASH result is averaged over 50 events per data point.}
  \label{fig:screening_mass_test}
\end{figure}

Figure \ref{fig:screening_mass_test} shows that the calculated gluon and quark screening mass sampled from SMASH fits perfectly the exact value calculated from the Boltzmann distribution function. It demonstrates that the number density corresponding to the range of temperature that we are interested in provides enough multiplicity for the momentum distribution function to form Boltzmann distribution.

\subsection{Equilibrium Stability}

As another benchmark check, it is important to verify that the medium is stable once it reaches an equilibrium state. This test also serves as a confirmation that the detailed balance principle is properly satisfied. Even though the detailed balance principle reinforces the same interparticle rate for back and forth directions, we still expect local short-term fluctuations where thermodynamic properties such as number density and partial energy slightly deviates from the supposed equilibrium. Therefore, stability will only be judged based on the long-term global behavior instead of focusing on these short-term local fluctuations. 

As we assume that the system equilibrates toward the Boltzmann distribution, we test the stability of the equilibrated medium by time evolving a medium initialized already in the equilibrium state. In this case, we initialized the medium with the Boltzmann distribution function at $T$ = 450 MeV, which corresponds to energy density $e \sim$ 65 GeV/$\text{fm}^{3}$ inside a box with 5 fm length. The initial spatial distribution is sampled as uniform distribution inside the box while the initial momentum distribution is sampled directly from the Boltzmann distribution. The simulation then runs for 100 fm to ensure that the equilibrium is stable over a long enough period of time given that equilibration based on parton cascade models often take about ~10 fm to achieve in the heavy-ion collision case \cite{Miller:2007ri, PhysRevLett.111.222302}. The following results are averaged over 50 events.

\begin{figure}[h]
  \includegraphics[width=\linewidth]{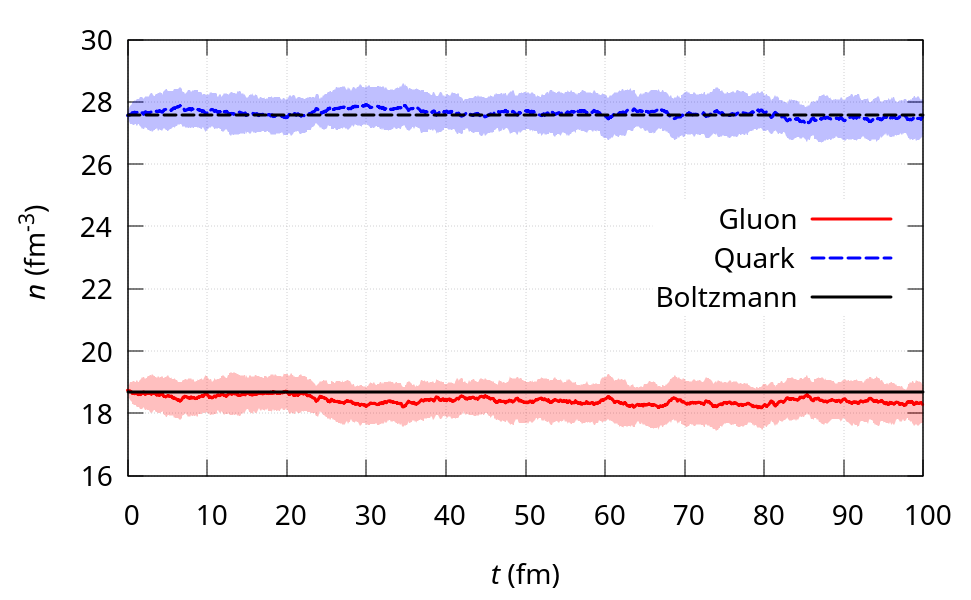}
  \caption{Parton number density stability test up to $t$ = 100 fm compared to the Boltzmann distribution at $T$ = 450 MeV given by the black line. Quark number density is given by blue color while gluon number density is given by red color. Statistical error is shown by the shaded area.}
  \label{fig:stability_test_number}
\end{figure}

\begin{figure}[h]
  \includegraphics[width=\linewidth]{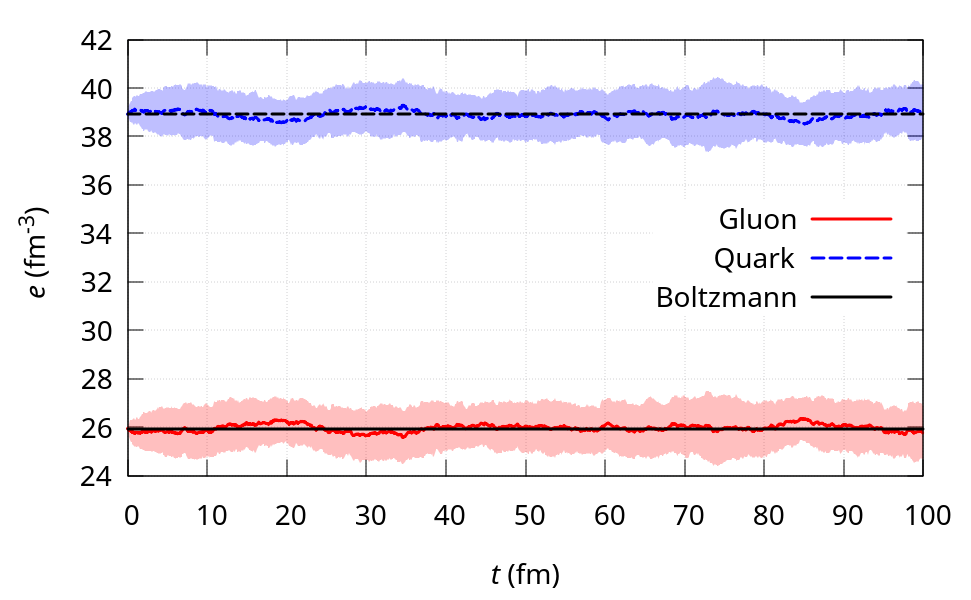}
  \caption{Energy density stability test up to $t$ = 100 fm compared to Boltzmann distribution at $T$ = 450 MeV given by the black line. Quark energy density is given by blue color while gluon energy density is given by red color. Statistical error is shown by the shaded area.}
  \label{fig:stability_test_energy}
\end{figure}

Figure \ref{fig:stability_test_number} shows that parton number density is indeed stable across the time evolution when the medium is initialized already in an equilibrium state. Moreover, despite the short-term fluctuation, the relatively stable number density for both quark and gluon also signifies that the system properly satisfies the detailed balance principle when it is in an equilibrium state. Similarly, Fig.~\ref{fig:stability_test_energy} demonstrates that energy density exhibits a similar behavior as well where both quark and gluon energy density are stable across the time evolution despite of the short-term fluctuation. The stable energy density and number density for both quark and gluon species guarantee that once the system reaches an equilibrium state, it will stay in the equilibrium for a long enough time scale compared to the typical kinetic theory simulation time scale in high-energy heavy-ion collisions. The stable number density and energy density that we have observed implies that the energy distribution function does not veer off from the initial Boltzmann distribution function, thus, the thermal equilibrium is also preserved once the system achieves an equilibrium state. Our next focus is to see whether a system initialized in a far-from-equilibrium state of the initial stage in the heavy-ion collision converges toward the same equilibrium state.

\section{Box Simulation Equilibration and Hydrodynamization}\label{sec:box_equilibration}

In this section, we start the medium in a far-from-equilibrium state in a box with a fixed size and periodic boundary condition to confirm whether the medium equilibrates toward the Boltzmann distribution and if the equilibrium state is stable or not once equilibrium is achieved. This setup ensures that the energy inside the box is fixed across the time evolution, thus simplifying the task of determining the temperature, energy density, and the number density of the system when equilibrium is reached. The goal of this section is to establish the baseline to determine when thermal and chemical equilibrium, as well as hydrodynamization, are achieved before applying the same definition in the expanding medium case.

\subsection{Initial Condition}

In contrast to the previous section, we prepare the initial condition in high-energy heavy-ion collisions determined using the mini-jet model, which produces a highly anisotropic far-from-equilibrium state. First, we calculate the number of nucleon-nucleon collisions in a single event using the Glauber Model based on the overlap of the nucleus $n_a$ and nucleus $n_b$,
\begin{equation}\label{eq:glauber_eq}
    N = 4 \nu \sigma_{\text{jet}} \int d^{2}x_{T}\ dz\ dt\ n_{\text{A}}(x_{T} + b, z - vt) n_{\text{B}}(x_{T} - b, z + vt).
\end{equation}
The factor 4 in the front is due to the assumption that each collision produces 2 free partons, and another extra factor of 2 is added to compensate for the higher order perturbation contribution due to we only consider up to the leading order in pQCD to calculate $\sigma_{\text{jet}}$. The jet cross-section is then given by 
\begin{equation}
    \sigma_{jet} = \int_{p^{2}_{\text{cutoff}}}^{4p^{2}_{\text{COM}}} dt \frac{d\sigma}{dt},
\end{equation}
where $p_{\text{cutoff}}$ is fixed to 2 GeV. We assume that if the momentum exchange in the collision is less than the cut-off, the resulting partons do not carry enough transversal momentum to break free from the nucleon. The differential cross section is calculated based on the pQCD up until leading order given by Eq.~(\ref{eq:full_cross_2_to_2}). The distribution of the nucleon inside the nucleus $n_{A}$ and $n_{B}$ is determined using Lorentz-boosted Wood-Saxon distribution function given by
\begin{equation}
    n(x_{T}, z) = \frac{\gamma n_{0}}{1+e^{\frac{\sqrt{x^{2}_{T}+(\gamma z)^{2}} - R}{d}}},
\end{equation}
where $d$ = 0.54 fm, $R$ = 1.12$A^{\frac{1}{3}}$ - 0.86$A^{-\frac{1}{3}}$, and $n_{0}$ is normalization constant such that the full spatial integral of $n$ will give the nuclei's atomic mass number \cite{Miller:2007ri}.

The initial spatial distribution of the partons is determined by sampling the overlap function used in the Glauber model calculation,
\begin{equation}
    \frac{dN}{dz d^{2}x_{T}} = \int dt\ n_{\text{A}}(x_{T} + b, z - vt) n_{\text{B}}(x_{T} - b, z + vt).
\end{equation}

After determining the number and the spatial distribution of the nucleon-nucleon collision, we treat each nucleon-nucleon collision as deep inelastic scattering, which determines the initial momentum distribution of the partons following the deep inelastic scattering differential cross-section,
\begin{equation}\label{eq:hic_moment_distb}
    \frac{d\sigma}{dp^{2}_{T}dy_{1}dy_{2}} = K \sum_{a,b} x_{1}f_{a}(x_{1}, p^{2}_{T}) x_{2}f_{b}(x_{2}, p^{2}_{T}) \frac{d \sigma_{ab}}{dt}.
\end{equation}
Here, $K$ is a coefficient fixed to 2 used to compensate for the small cross-section value since we only calculate up to the tree-level diagram. In our calculation, we used the nuclear neural-network partonic distribution function (nNNPDF) \cite{AbdulKhalek:2019mzd} for the partonic distribution functions $f_{a}$ and $f_{b}$.

A sanity check is done by calculating the total cross section of the nucleon-nucleon collision including all 2-to-2 collision channels, by setting $p_{\text{cutoff}}$ = 1.8 GeV following PYTHIA's prescription for Au-Au collision at $\sqrt{s_{NN}} =$ 200 GeV \cite{Bierlich2018}, and fixing coupling constant $\alpha_{s}$ = 0.3.
\begin{equation}
    \sigma_{NN} = \int dx_{1} dx_{2} dt \sum_{\text{channels}} x_{1}f_{1}(x_{1}) x_{2}f_{2}(x_{2}) \frac{\sigma}{dt} = 44.57 ~\text{mb}.
\end{equation}
Compared to COMPETE prediction of 51.79 mb \cite{PhysRevLett.89.201801} and STAR experiment of 54.67 mb \cite{2020135663}, our calculation gives $\sim$10\% lower cross section. This difference is credited to our cross-section calculation, which is only considered up to the leading order. 

Under this setup, we start the system with a uniform spatial distribution inside the box and the momentum distribution mimicking the initial momentum distribution in the high-energy heavy-ion collision case as given in Eq.~(\ref{eq:hic_moment_distb}). In this part, we focus on Au-Au at $\sqrt{s_{NN}} =$ 200 GeV collision system and limit the momentum rapidity of the initial partons to $|y| < 0.5$. The initial number of partons is determined by Eq.~(\ref{eq:glauber_eq}) which yields 250 nucleon-nucleon collisions. As each collision is assumed to produce 2 free partons, this corresponds to 500 initial partons. Then, as the initial state in the heavy-ion collision is gluon-dominated, we fixed the species of all 500 initial partons as gluons. The average energy of each gluon based on the aforementioned formulation is found to be $\langle E_{\text{parton}} \rangle$ = 3.25 GeV, thus the average total energy of the system is $\langle E_{\text{Total}} \rangle$ = 1629 GeV. Because the box has periodic boundary condition, the total energy of the system is preserved across the time evolution. As the equilibrium state is defined by the Boltzmann distribution, this total energy of the system corresponds to $T$ = 0.447 GeV as here we consider only 2 quark species, the up quark and down quark. This temperature is the assumed temperature of the system that we should get when the system has fully equilibrated.

One of the main distinctions of our model is the addition of gluon absorption and radiation when interacting with quarks. Previous models often only consider gluon absorption and radiation either through decay or 2-to-3 interactions exclusively to gluons since the initial state of the system is gluon-dominated \cite{PhysRevLett.114.182301, PhysRevC.48.1275}. In contrast to the gluon-dominated system in the initial state, the quark number density in the equilibrium state is higher than the gluon number density. Hence, the transition from a gluon-dominated into a quark-dominated system should happen before the system reaches an equilibrium state. In fact, we found that this switch occurs much earlier before the system achieves equilibrium. Thus, the gluon absorption and radiation through the interaction with quarks plays an important role in the equilibration process.

We start by simulating the system with and without the extra channels qg → qgg and qq → qqg with 50 events for each case in a fixed volume box with the length of 5 fm. Then, in order to establish conditions to identify equilibration and hydrodynamization, we will evolve the medium for 50 fm, which is far longer than the typical QGP equilibration time in kinetic theory, to ensure the system is equilibrated and hydrodynamized by the end of its time evolution.

\subsection{Chemical and Thermal Equilibration}

The concept of equilibrium state in statistical physics refers to the state in which the entropy is maximized where entropy is given by Shannon's entropy

\begin{equation}
    S = -\int_{\Omega}f(E)\text{ln}(f(E))dE
\end{equation}
across all the possible energy states $\Omega$. The common approach to maximize this entropy follows the Lagrange multiplier, which eventually results in a probability distribution function

\begin{equation} \label{eq:distrib}
    f(E)=Ae^{-BE}
\end{equation}
with constants A and B. For a classical ultra-relativistic ideal gas in equilibrium, the energy distribution follows

\begin{equation}
    \frac{dN}{NdE} \propto \frac{1}{2T}e^{-\frac{E}{T}}.
\end{equation}
Following from Eq.~(\ref{eq:distrib}), this distribution function is guaranteed to maximize entropy, then in turn, guarantees equilibrium state. Hence, we will use this as our baseline to define the equilibrium state throughout our analysis.

As we have established that the Boltzmann distribution guarantees equilibrium state, we will then use number density as a proxy to identify chemical equilibrium,

\begin{equation}
    n = \int \frac{d^{3}p}{(2\pi)^{3}} f(E).
\end{equation}
Substituting $f(E)$ with the Boltzmann distribution yields

\begin{equation}
    n = \frac{g_{\nu}}{\pi^{2}}T^{3},
\end{equation}
where $g_{\nu}$ is the degeneracy factor with gluon having $g_{\nu}^{\text{gluon}}$ = 8 colors $\times$ 2 spin and $g_{\nu}^{\text{quark}}$ = 3 colors $\times$ 2 spin $\times$ 2 quark-antiquark $\times$ 2 species as we only consider up and down quarks.

\begin{figure}[h]
  \includegraphics[width=\linewidth]{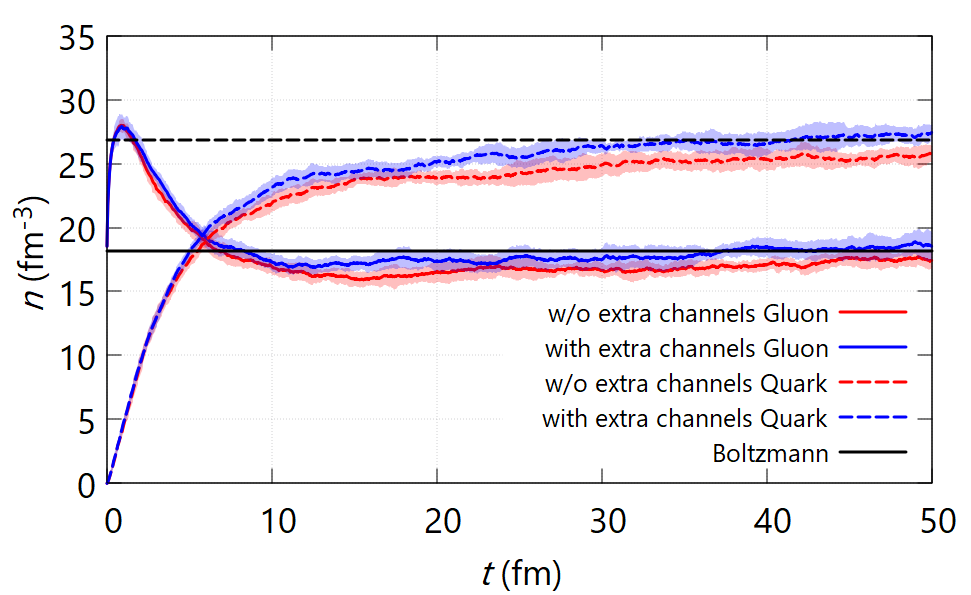}
  \caption{Parton number density evolution up to 50 fm. Quark and gluon show convergence toward the Boltzmann limit represented by the black dashed line and black solid line, respectively. The extra channels case colored in blue shows a faster convergence compared to the case without extra channels.}
  \label{fig:box_equilibration_parton_number}
\end{figure}

Figure \ref{fig:box_equilibration_parton_number} shows the evolution of quark and gluon number density separately both with and without the extra channels. We see that the transition from a gluon-dominated system into a quark-dominated system happens between 5-10 fm. We also see that the extra channels hasten the saturation toward the Boltzmann limit both for quark and gluon number density with gluon showing a clearer difference as it reaches the Boltzmann limit around 30 fm with the extra channel case, whereas the case without the extra channel has not reached Boltzmann limit even after 50 fm even though it does seem to be approaching the Boltzmann limit as well.

\begin{figure}[h]
  \includegraphics[width=\linewidth]{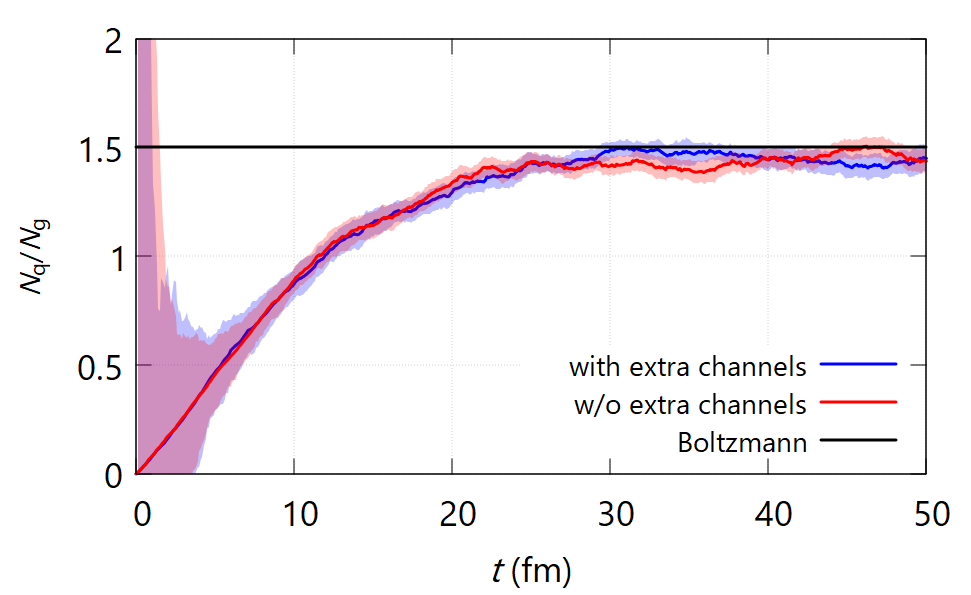}
  \caption{Parton number ratio evolution up to 50 fm. Despite the faster convergence for the extra channels case in terms of number density, we see a similar convergence rate for the parton number ratio in both with and without extra channels cases.}
  \label{fig:box_equilibration_parton_number_ratio}
\end{figure}

Another indicator of chemical equilibration is the number-density ratio which is determined by the distribution function and the degree of freedom of each particle species. Given that we assume Boltzmann distribution for both quark and gluon, thus the ratio between quark and gluon solely depends on their degree of freedom. Figure \ref{fig:box_equilibration_parton_number_ratio} shows that despite the difference in the number density saturation rate, the parton number ratio showed similar time evolution for both cases. The saturation behavior seen at the later time also demonstrates that the system satisfies the detailed balance principle, which is an important condition for chemical equilibrium.  We see that the addition of the extra channels changes the parton number convergence rate toward the Boltzmann limit, in contrast, such an addition did not change parton number ratio convergence rate toward the Boltzmann limit, it implies that the quark-antiquark pair production and annihilation achieve the detailed balance principle at a similar time scale regardless of the extra channels, while the gluon production and absorption reach the detailed balance principle faster with the extra channels. This makes perfect sense as the extra channels are the gluon production and absorption due to the quark-quark or quark-gluon interaction, which also indicates that the effect of these channels on equilibration and thermalization is not negligible.

Regarding the thermal equilibrium, unfortunately, it is difficult to define temperature for a non-equilibrium case due to the fact that we do not have a well-defined distribution function for a non-equilibrium state. Hence, we adopt a top-down approach by first affirming the assumption that the equilibrium state is defined by the Boltzmann distribution function before extracting the temperature based on this premise.

\begin{figure}[h]
  \includegraphics[width=\linewidth]{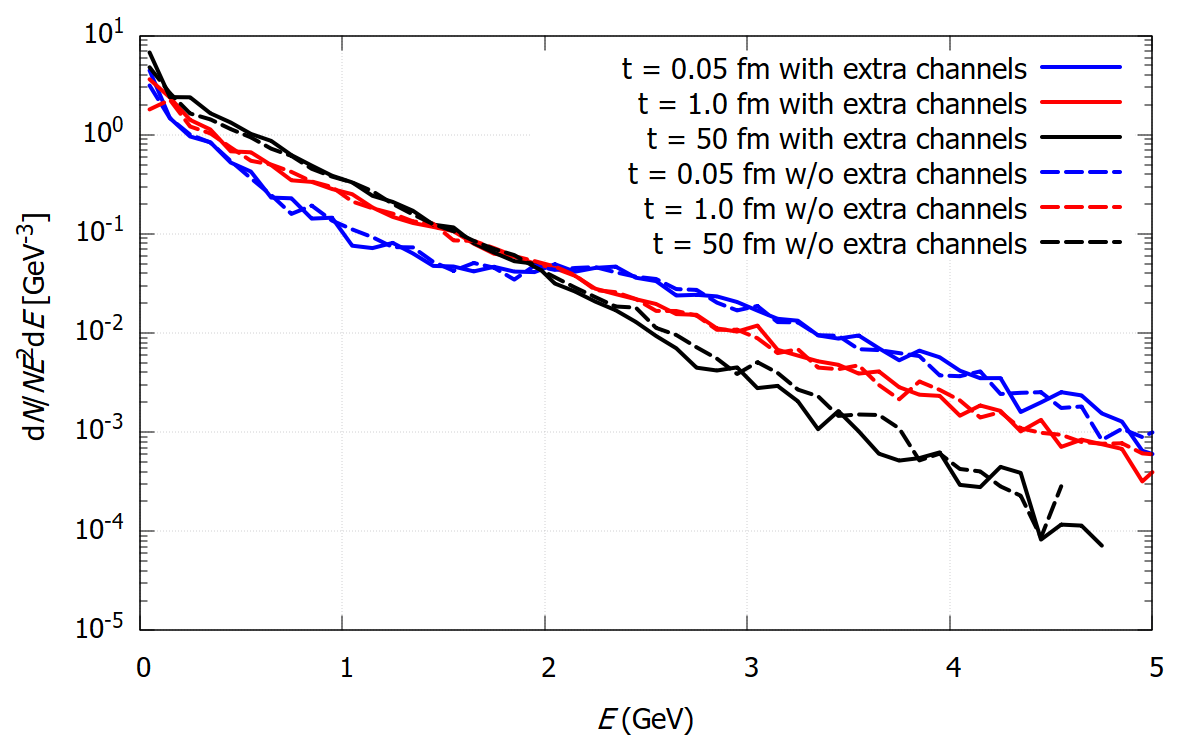}
  \caption{Energy spectrum across different times ($t$ = 0.05 fm, 1 fm, 50 fm from the top to the bottom histogram, respectively) for the case with the extra channel shown by the solid line and the case without the extra channel shown by the dashed line.}
  \label{fig:box_equilibration_energy_spectrum}
\end{figure}

Figure \ref{fig:box_equilibration_energy_spectrum} proves that the energy spectrum of the medium follows a negative exponential function at the later time both for with and without the extra channel cases. It confirms our assumption that the energy spectrum converges toward the Boltzmann distribution function which is already satisfied after 1 fm. The slope of the energy spectrum continues to grow steeper with time, which describes the time evolution of the temperature as the system converges toward the equilibrium state. Moreover, in contrast to the clear difference in the number density, we see no significant difference in the energy spectrum between with the extra channel case and without the extra channel case.

Although we can apply linear regression to extract temperature information from the energy distribution function at this stage, we shall first look at homoskedasticity and $\text{R}^{2} \text{-value}$ from the linear regression. 
Homoskedasticity is a measure of how close the difference between the data and the mathematical model resembles gaussian white noise, while $\text{R}^{2} \text{-value}$ measures how well the mathematical model fits the data based on explained variance. A more detailed explanation regarding these two values is given in the appendix~\ref{appendix:r2_homosked_explanation}. This test will serve to quantify how close the energy distribution function is to the Boltzmann distribution function, thus validating whether we can extract temperature by assuming the Boltzmann distribution function.

In our calculation, homoskedasticity will be measured using the Breusch-Pagan test in which for 1 free parameter case, homoskedasticity is consistent when the measured $\chi < \text{3.841}$. Not only this confirms the homoskedasticity of the data, which is the prerequisite for the validity of the linear regression, but it also serves as an indicator of how close the energy spectrum resembles the Boltzmann distribution function.

A closer examination of the Boltzmann distribution function fits against the energy spectrum discussed in the appendix indicates that the upper limit of linearity in the log plot of the energy spectrum is found to be at 3 GeV. This is consistent with the upper limit often used in the hydrodynamic condition where its constituent particles come from the soft region in the energy spectrum. As this method has shown consistency and ability to precisely identify the upper limit in the regression approach, it is a promising approach that might be applied in different cases as well such as in the effective modeling and data analysis of experimental data.

\begin{figure}[h]
  \includegraphics[width=\linewidth]{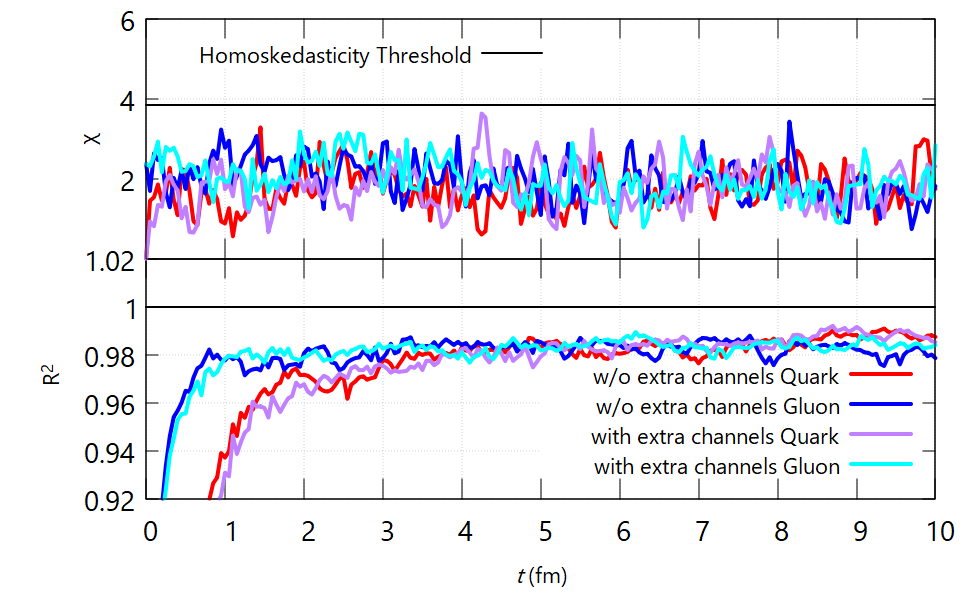}
  \caption{Linear regression fit up to $t$ = 10 fm. The quality of fit is represented by $\chi$ (top) and $\text{R}^{2}$-value (bottom) across the time evolution. The energy range used in the fit is fixed between 0 GeV and 3 GeV. The convergence toward the Boltzmann distribution can be clearly seen by the $\text{R}^{2}$-value saturation toward 1.}
  \label{fig:box_equilibration_regression_fit}
\end{figure}

The linear regression fit test for the first 10 fm of the evolution as shown in Fig.~\ref{fig:box_equilibration_regression_fit} proved that the linear regression result is invalid for the first 1 fm for gluon and up to 3 fm for quark. We also observe the $R^{2}$ which describes how close the data resembles a linear function exhibits convergence behavior toward the same value while with and without extra channels do not show a significant difference. However, we clearly see that gluon converges much earlier at 1 fm in contrast to quark which takes until 6 fm to saturate. This difference in thermalization timescale based on species was also observed with the same tendency in previous models as well \cite{Sengupta1997}.

\begin{figure}[h]
  \includegraphics[width=\linewidth]{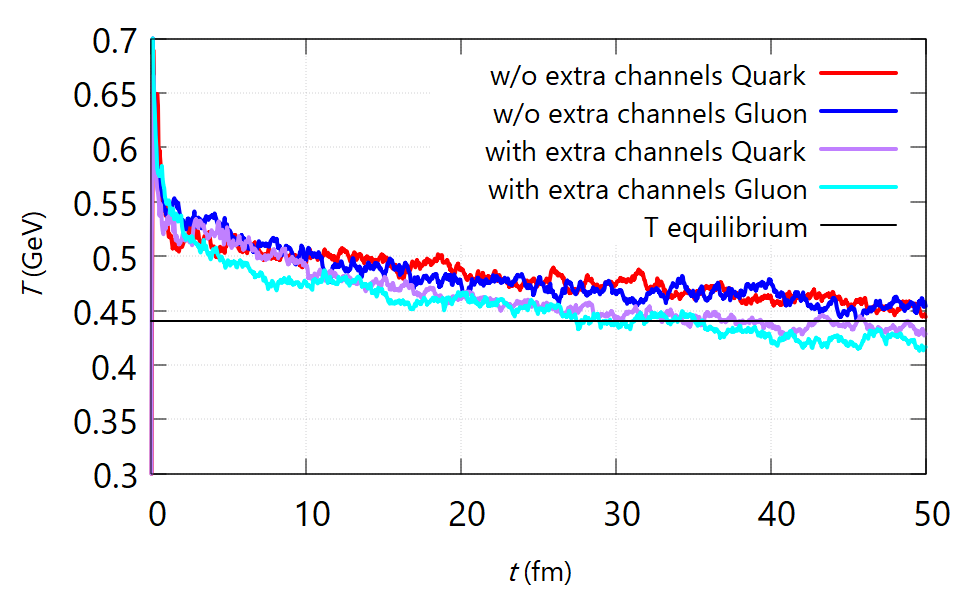}
  \caption{Temperature evolution across the time evolution up to $t$ = 50 fm. $T_{\text{equilibrium}}$ is fixed at 0.45 GeV which corresponds to the total energy inside the box.}
  \label{fig:box_equilibration_temperature}
\end{figure}

As we have confirmed that linear regression is valid and fit the model after 3 fm, we then extract the temperature of the system as shown in Fig.~\ref{fig:box_equilibration_temperature}. Even though we see both cases and species measure higher temperature than the corresponding equilibrium temperature based on the total energy of the system, we see clearly that with extra channels case shows a faster approach toward equilibrium temperature both for quark and gluon at between 30-40 fm time mark, while without extra channels took full 50 fm to match the corresponding equilibrium temperature. This reinforces the idea that the quark interaction does play an important role as well in the equilibration process.

\subsection{Momentum Anisotropy}
Momentum anisotropy is one of the indicators of hydrodynamization. We measure the anisotropies of the system as
\begin{equation}
    \frac{2 \langle P_{z}^{2} \rangle}{\langle P_{x}^{2} + P_{y}^{2}\rangle} = \frac{2 \langle P_{z}^{2} \rangle}{\langle P_{T}^{2} \rangle},
\end{equation}
which is calculated separately for gluon and quark and averaged over all particles for each event. Because isotropy requires $P_{x} = P_{y} = P_{z}$, thus the above value is equal to 1 when isotropy is reached.

\begin{figure}[h]
  \includegraphics[width=\linewidth]{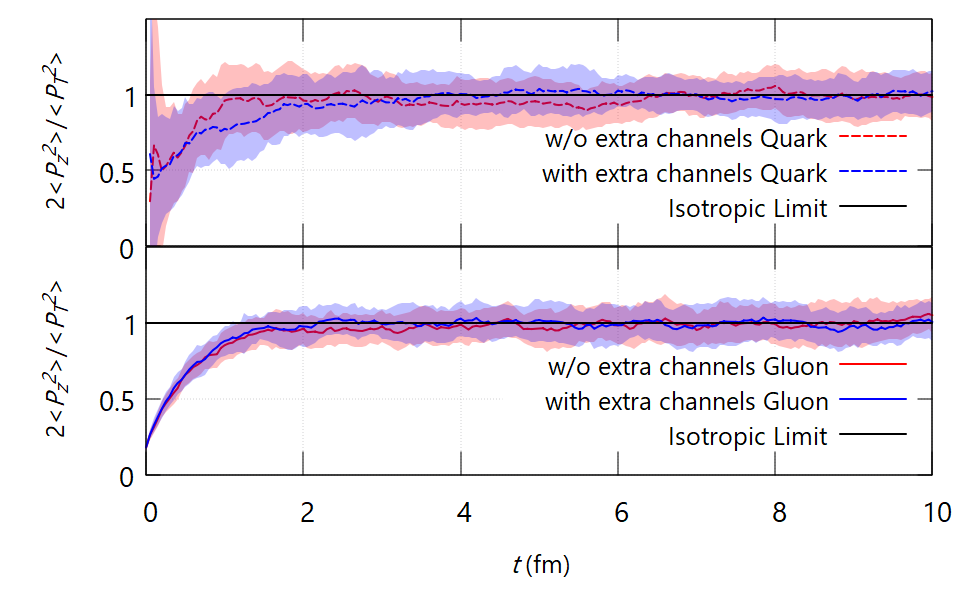}
  \caption{Momentum anisotropy measured across the time evolution for quark (top) and gluon (bottom). The system rapidly isotropizes with a similar rate both for with and without the extra channels cases.}
  \label{fig:box_equilibration_anisotropy}
\end{figure}

Figure \ref{fig:box_equilibration_anisotropy} shows that momentum isotropy is already achieved as soon as 2 fm for both with and without extra channel cases. The anisotropy results suggest that the medium reaches momentum isotropy well before achieving either global thermal or chemical equilibrium. However, as hydrodynamic calculations assume local equilibrium is already established, it might be insufficient to define hydrodynamization solely based on the momentum anisotropy.  There is a need to either find a better identifier of hydrodynamization or to instead define hydrodynamization based on several conditions. Such an approach is left for future work as the main goal of this paper is to show that the medium indeed reaches momentum isotropy after given enough time.

Up until this point, we have confirmed the system does properly reach thermal and chemical equilibration as well as momentum isotropy starting from an anisotropic far-from-equilibrium state for a box simulation with periodic boundary condition. We also have seen that in this case, momentum isotropization is reached first at $t \sim$ 2 fm, followed by thermal equilibration at $t \sim$ 3 fm, and finally chemical equilibrium at a much later time scale at $t \sim$ 30 fm.

\section{Expanding Medium Equilibration and Hydrodynamization}\label{sec:hic_equilibration}
In this section, we will examine the time evolution for an expanding medium created in the high-energy heavy-ion collision. Moreover, as we no longer limit the momentum rapidity at $|y| <$ 0.5 and the initial spatial distribution is decided by the Glauber model instead of a uniform distribution, we will have a highly anisotropic medium with a strong longitudinal expansion and dense medium at the beginning of the simulation compared to the box simulation.

\subsection{Chemical and Thermal Equilibration}
Due to the nature of an expanding medium, the characteristics of the medium formed in the peripheral region is different compared to the central region. As the central region is the area with the highest particle density and temperature, it is reasonable to assume that equilibrium is more likely to be reached in the central region with enough particle multiplicity. Thus we limit our analysis to the central region in the form of a cylinder with expanding length defined by spatial rapidity $\eta \in [-0.5, 0.5]$ and fixed radius based on nuclear effective radius given by $R$ = 1.12$A^{\frac{1}{3}}$ - 0.86$A^{-\frac{1}{3}}$ \cite{Miller:2007ri}. Mass number $A$ = 197 for Au-Au collision, hence $R$ = 6.4 fm. In this section, we only focus on the central collision with impact parameter fixed at $b$ = 0 fm.

\begin{figure}[h]
  \includegraphics[width=\linewidth]{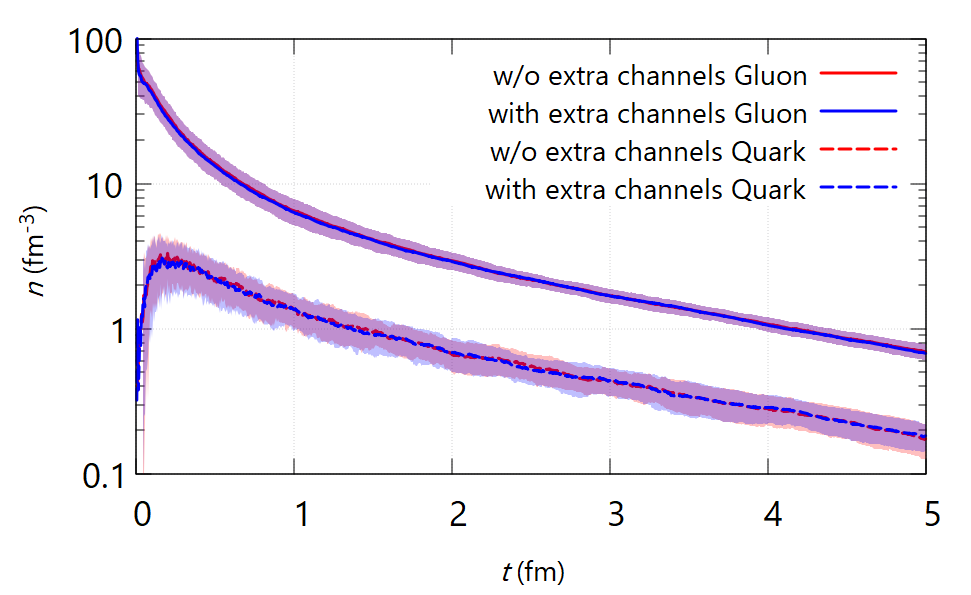}
  \caption{Number density evolution up to 5 fm for quark (dashed line) and gluon (solid line) in the mid-rapidity region $\eta \in [-0.5, 0.5]$. With extra channels case (red) is identical compared to without extra channels case (blue).}
  \label{fig:expanding_equilibration_number_density}
\end{figure}

Applying the same approach that we employed in the box equilibration case, we first examine the parton number density evolution of the system shown in Fig.~\ref{fig:expanding_equilibration_number_density}. Naturally, in an expanding medium, the density of the central region decreases with time; thus, it is clear that one cannot rely on saturation behavior in the parton number density to identify chemical equilibration. In this case, we identify chemical equilibration by comparing measured parton number density against the Boltzmann limit for a given effective temperature. As such, we will change our approach by first verifying whether thermalization is properly reached before examining chemical equilibration, where we shall argue that we have a well-defined effective temperature.

Moreover, in contrast with the box simulation case, here we do not see the switch between the gluon-dominated system into quark-dominated system despite the strong production of quarks in the beginning. However, we see the same consistency of the similarity between the with and without extra channel case as this tendency is also observed in the first 10 fm of the time evolution in the box simulation case.

\begin{figure}[h]
\includegraphics[width=\linewidth]{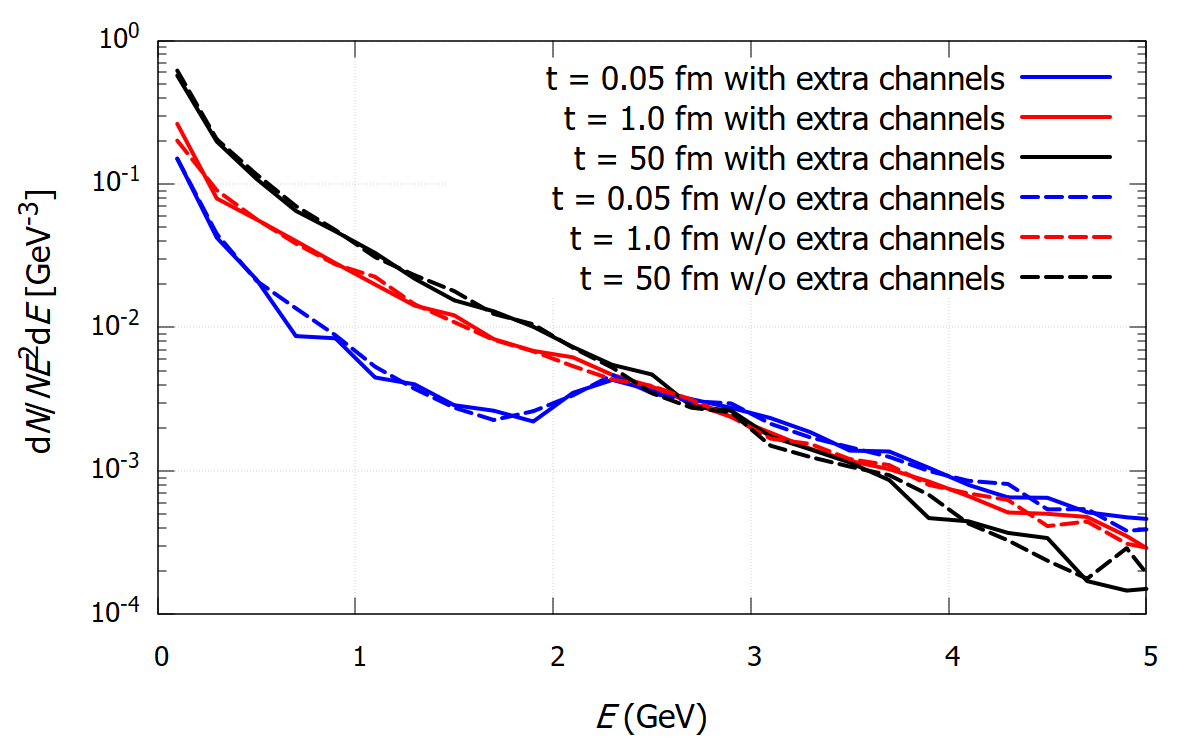}
  \caption{Energy spectrum for $t$ = 0.005 fm, 0.1 fm, and 5 fm. The extra channels case is shown with the solid line, while without extra channels case is shown with the dashed line. Lack of low-energy gluon in the early time is due to the initialized gluons cut-off fixed at $p_t>2\text{ GeV}$.}
  \label{fig:expanding_equilibration_espect}
\end{figure}

Figure \ref{fig:expanding_equilibration_espect} shows that the log of the energy spectra converges into a linear function, hence we shall apply the linear regression fit to find the slope of the spectra which will define the effective temperature of the medium. We assume a similar energy spectrum's convergence behavior as in the box simulation and use the same energy region that we have identified to have the best fit with the linear function in the last section, from 0 to 3 GeV as shown in the appendix~\ref{appendix:r2_homosked_explanation}.  But before we extract the temperature, we shall again apply linear regression fit analysis to determine the region where we can trust the fitting result.

\begin{figure}[h]
  \includegraphics[width=\linewidth]{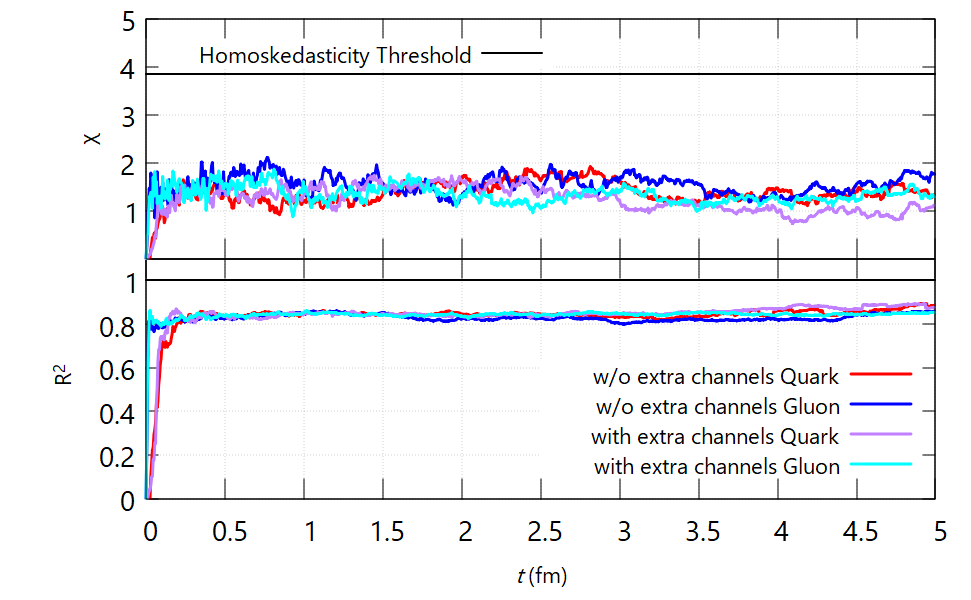}
  \caption{Linear regression fit analysis from energy spectrum focusing on $\eta \in [-0.5, 0.5]$ and $E \in [0, 3] \text{ GeV}$. $\text{R}^{2}$-value (bottom) shows rapid convergence toward the Boltzmann distribution with quark (red and purple) converging later than gluon (blue and teal).}
  \label{fig:expanding_equilibration_regression_fit}
\end{figure}

Similar to the box equilibration case, we see from Fig.~\ref{fig:expanding_equilibration_regression_fit} that $\text{R}^{2} \text{-value}$ for quark and gluon both saturate very early during the time evolution ($\sim 0.2$ fm) and slowly increase toward 1 at the later time. There is also a clear distinction between quark and gluon where gluon saturates very early during the time evolution, while quark is slightly lagging behind. On the other hand, homoskedasticity is also satisfied throughout the time evolution. The slow increases in the $\text{R}^{2} \text{-value}$ at the later time can be understood as the energy spectrum of the particles fits the Boltzmann distribution function better at the later time as the equilibration continues. Judging from the behavior of the $\text{R}^{2} \text{-value}$, it seems that the medium thermalizes very early at $t \sim$ 0.2 fm which is much sooner than the box equilibration counterpart, this is due to the extremely high density and multiplicity region formed in the central region at the very early time which promotes interparticle interaction and hence thermalization. This result is a relatively fast equilibration compared with various previous papers that suggested a rapid thermalization time for heavy ion collision lies around 0.5-2 fm \cite{BAIER200151, Xu:2004mz, PhysRevD.89.074011}. However, whether this thermal equilibrium is broken or not at the later time signalled by the lowering of $\text{R}^{2} \text{-value}$ is another matter that needs further examination and discussion.

In contrast to the box simulation case shown in Fig.~\ref{fig:box_equilibration_regression_fit}, the $\text{R}^{2}$-value in the expanding medium from Fig.~\ref{fig:expanding_equilibration_regression_fit} exhibits faster saturation behavior. This is due to the highly dense matter in the expanding medium where the measured parton density reaches 3 times higher density than in the box simulation. This causes a higher interaction rate, which in turn drives a faster thermal equilibration time than in the box simulation.

\begin{figure}[h]
  \includegraphics[width=\linewidth]{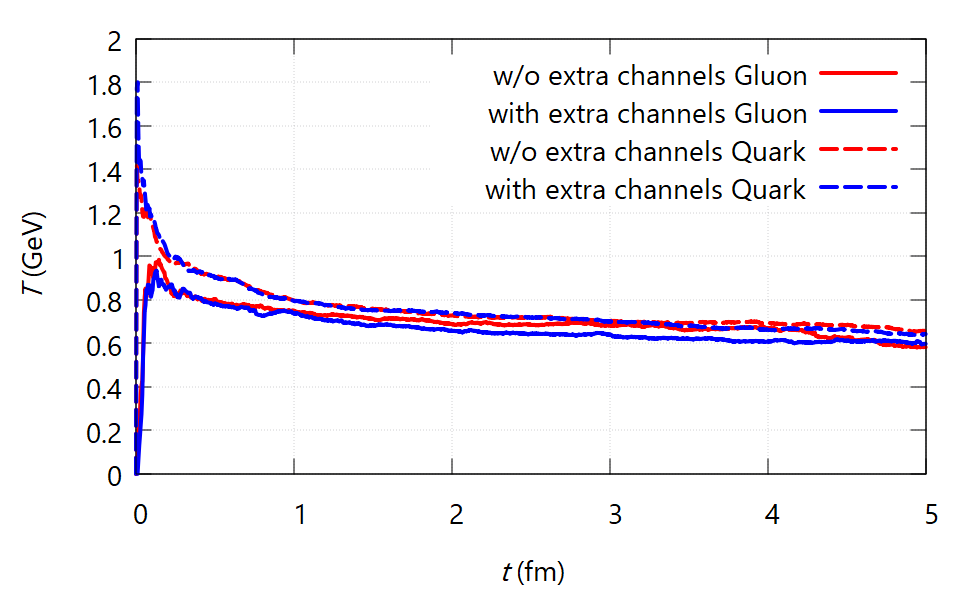}
  \caption{Temperature evolution up to 5 fm. Quarks (dashed line) show a different temperature profile compared to gluons (solid line) at the beginning of the time evolution and converge at a later time. We see almost identical time evolution for both with extra channels (red) and without extra channels (blue).}
  \label{fig:expanding_equilibration_temperature}
\end{figure}

Figure \ref{fig:expanding_equilibration_temperature} shows the temperature extracted using a linear regression fit from the energy spectrum. The temperature in an expanding medium is understandably constantly decreasing as the medium continues to expand. Quark saturates at a slower rate compared to gluon because gluon interaction rate is higher than quark which in turn drives a faster convergence toward equilibrium, while the rapid fluctuation at the early time for both quark and gluon is due to the poor fit of linear regression which we have observed as well in Fig.~\ref{fig:expanding_equilibration_regression_fit}. 

As we have extracted the effective temperature of the medium, we identify chemical equilibration based on fugacity which is defined as
\begin{equation}
    \nu_{q/g}(T_{q/g}(t)) = \frac{N_{q/g}(t)}{N^{\text{Boltzmann}}_{q/g}(T_{q/g}(t))}.
\end{equation}
Fugacity effectively measures how close the current parton number density is to the parton number density calculated using the Boltzmann distribution function.

\begin{figure}[h]
  \includegraphics[width=\linewidth]{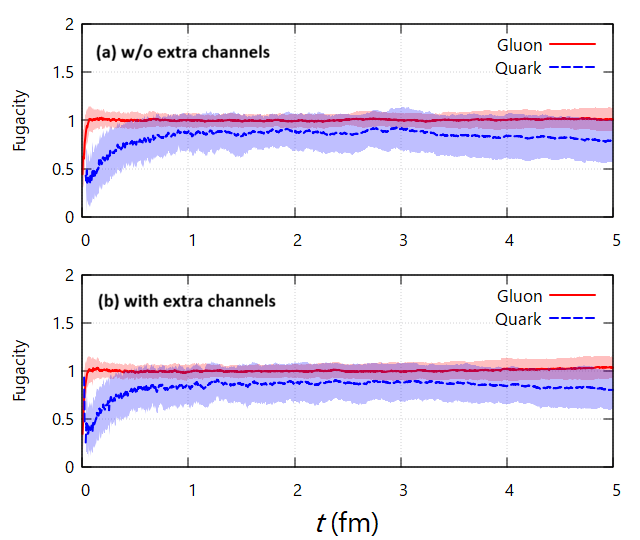}
  \caption{Fugacity up to 5 fm. Gluon is shown in blue and quark is shown in red. Without extra channels (a) and with extra channels (b), the cases show a similar time evolution profile.}
  \label{fig:expanding_fugacity}
\end{figure}

The measured fugacity shown in Fig.~\ref{fig:expanding_fugacity} is to be expected as we see in Fig.~\ref{fig:expanding_equilibration_number_density} that number density rapidly decreases due to the expanding medium, but Fig.~\ref{fig:expanding_equilibration_temperature} shows a rather slow temperature decrease after the very early time, hence we see a rapid increase in fugacity in the early time followed by a slow decrease in fugacity at the later time. 

The contrast between the saturation time between quark and gluon can be explained since gluon started from a highly saturated state at high temperature which then rapidly cools down toward its respective temperature for its energy density, whereas we initialized the system with zero quark while the cross-section for quark-antiquark pair production is proportional to $\frac{1}{s^{2}}$, thus the quark production is more prominent in low s situation at the later time when temperature has somewhat cooled down, causing the increase of fugacity in quark to lag behind gluon. Moreover, because the switch between a gluon-dominated system to a quark-dominated system does not occur in the expanding medium as demonstrated by Fig~\ref{fig:expanding_equilibration_number_density}, we understandably find a larger error bar in quark compared to gluon.

\subsection{Momentum Anisotropy}

\begin{figure}[h]
  \includegraphics[width=\linewidth]{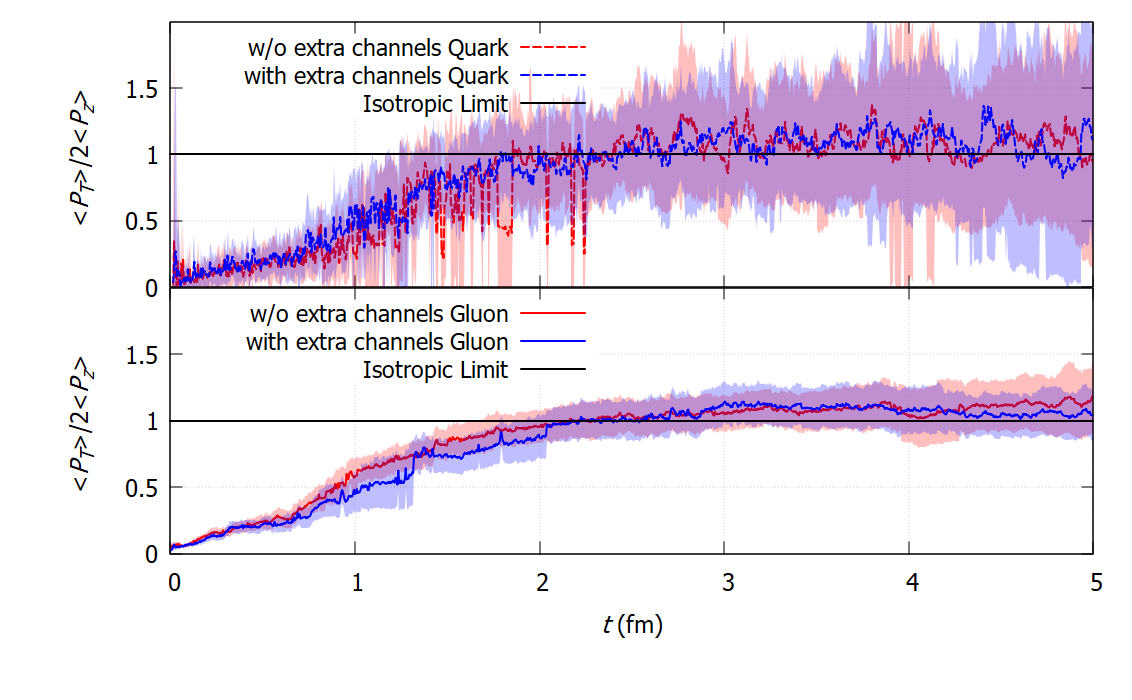}
  \caption{Comparison of momentum anisotropy between quark (top) and gluon (bottom). Both cases reach isotropy around the same time with the quark showing a stronger fluctuation due to the lower particle number.}
  \label{fig:expanding_anisotropy}
\end{figure}

In contrast to the box simulation with the periodic boundary condition, the non-thermalized particles in the expanding medium case with a large momentum at the collision axis tend to move away and leave the bulk of the medium. As these jets are mixed between the thermalized particles in the central rapidity region at the beginning of the time evolution, naturally we will have stronger longitudinal mean momentum than mean transversal momentum. Hence, the same definition of anisotropy in the box simulation case ($2 \langle P^{2}_{z} \rangle / \langle P^{2}_{T} \rangle$) will approach infinity at $t \sim 0$; thus, in the expanding medium case, we will use the inverse of this definition of anisotropy.

Figure \ref{fig:expanding_anisotropy} shows the first 5 fm evolution of the momentum anisotropy. It shows a similar anisotropy evolution between quarks and gluons. Isotropy is reached around 2 fm where quarks show a fluctuating value but still around the isotropic line, while gluon shows a stronger transverse momentum momentarily before returning to the isotropic line at a later time.

This result is considerably longer than the regular initialization time in hydrodynamic simulation for Au-Au collision at RHIC energy of ~1.0 fm \cite{PhysRevC.99.044902, Gotz:2025wnv, Chaudhuri:2013yna}, while anisotropic hydrodynamic model even suggests faster initialization time around ~0.2 fm \cite{Strickland:2024moq}. On the other hand, the rather slow hydrodynamization time is fairly consistent with the other kinetic theory study in hydrodynamization \cite{Rajagopal2025}, as indeed this problem is a long-standing problem in the kinetic theory approach. Moreover, we see no notable difference between the with and without extra channels cases. Since in particular for the Au-Au collision at RHIC energy, we see no switch between a gluon-dominated system to a quark-dominated system, hence the extra channels focused on quark interaction did not affect much of the result.

\subsection{Knudsen Number}

Knudsen number determines whether a system behaves closer to a discrete system of individual particles or a continuous fluid-like system, thus, it is often employed to measure the applicability of hydrodynamic description. Knudsen number is defined as
\begin{equation}
    K_{n} = \frac{\lambda_{\text{mfp}}}{l},
\end{equation}
which describes the ratio between the mean free path and the characteristic length of the system. Hence, when $K_{n} \geq 1$, a single particle can travel from one end of the system to the other end without going through a single interaction which will equate to a free propagating system. While $K_{n} \leq 1$ suggests a strongly interacting system which indicates hydrodynamic system.

As the Knudsen number is proportional to the inverse of the characteristic length, it plays an important role in determining the Knudsen number. Unfortunately, there is no single accepted method to define the characteristic length of the medium. The common approach in hydrodynamic model is based on the local gradient of fluid dynamical quantities such as the energy density or pressure \cite{Denicol:2012cn, Strickland:2017kux}, but such a definition cannot be used in point-particle system of parton cascade model, on the other hand, defining the initial characteristic length as the initial size of the overlapping area in the collision and assuming it to expand relativistically with the speed of light \cite{Nugara:2024net} is rather overly simplistic since it disregards local information. As such, before we can have a meaningful discussion regarding the Knudsen number, it is important to closely scrutinize how to define the characteristic length of the medium. 

\subsubsection{Expectation-Maximization Characteristic Length}

We determine the characteristic length by assuming a single cluster of 3D Gaussian shape at the center point of the 2 colliding nuclei. The width of the gaussian is defined by the expectation-maximisation clustering algorithm where we set its gaussian width $\sigma$ to maximize the expectation value
\begin{equation}
    \mathcal{F} = \sum_{i}^{N}-\frac{1}{2}\frac{x_{i}^{2}}{\sigma^{2}} - \frac{1}{2} \text{ln}(\sigma^{2}),
\end{equation}
then the characteristic length of the medium is defined as 
\begin{equation}
    l=2 \langle r \rangle,
\end{equation}
where the mean radius of the cluster is calculated as
\begin{equation}
    \langle r \rangle=\sqrt{\sigma_{x}^{2} + \sigma_{y}^{2} + \sigma_{z}^{2}}.
\end{equation}

\begin{figure}[h]
  \includegraphics[width=\linewidth]{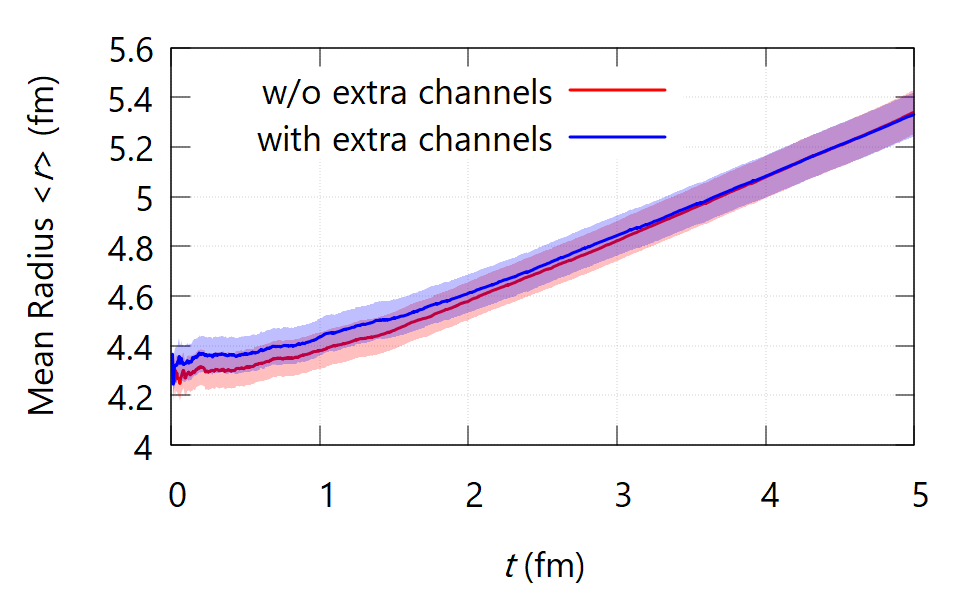}
  \caption{Mean cluster radius extracted from the spatial distribution of the particles across the time evolution, the value shown is to be expected considering that we only focus on the central collision and the Woods-Saxon radius of Au is around 6 fm.}
  \label{fig:mean_radius}
\end{figure}

As we have an expanding medium, naturally, the mean cluster radius is increasing with time. The measured value is also similar to the effective radius of gold nuclei calculated using the Wood-Saxon radius of ~6 fm. This is due to the fact that we only focus on ultra-central collisions with zero impact parameter. However, we see in Fig.~\ref{fig:mean_radius} that with extra channels case increases slightly slower than without extra channel case. More interactions mean slower expansion as particles are less likely to fly away from the medium without interaction.

Fitting per timestep and per event assures the characteristic length to adjust to the fluctuation of the shape of the medium. Moreover, compared to the relativistic expansion assumption where the speed of the expansion is the speed of light which produces a linear plot between characteristic length and time, here we see a slower expansion at the early time which speeds up at the later time, but still does not reach the speed of light. Bjorken expansion only assumes longitudinal expansion at the speed of light, hence the slower expansion is more realistic if we consider the slow expansion in the transverse direction.

In this case, we assumed that the medium is made of a single 3D Gaussian-like cluster located in the center of the collision whereas in practice, the energy distribution inside the single cluster may vary greatly and forming smaller hotspots of high energy density. Then each of these hotspots can arguably be considered as an individual hotspot with a smaller size scale. However, in our calculation, we opted to consider the collective of these hotspots as a single cluster because the energy exchange between these hotspots will not change the total energy of the whole cluster. In this case, the only important energy loss is from the particles travelling too far away from the center and hence outside of the main cluster.

\subsubsection{Shannon Entropy Maximization Characteristic Length}

Another way to define the characteristic length of the medium is by dividing the medium into smaller cells to isolate the smaller hotspots during the time evolution. In contrast to the previous approach, we did not make an assumption regarding the number of clusters. Moreover, the previous method defines the characteristic length based on the global shape of the medium and ignores local energy distribution, while this method calculates the characteristic length with the aim of maximizing entropy. Since maximum entropy indicates an equilibrium state, this method is more in tune with the local equilibrium assumption often used in hydrodynamic calculations. 

The cell size in this case is defined such that it maximizes the Shannon entropy calculated from the energy density spectrum of the cells given as
\begin{equation}
    S=-\sum_{i}\rho_{i}~\text{ln}(\rho_{i}).
\end{equation}

\begin{figure}[h]
  \includegraphics[width=\linewidth]{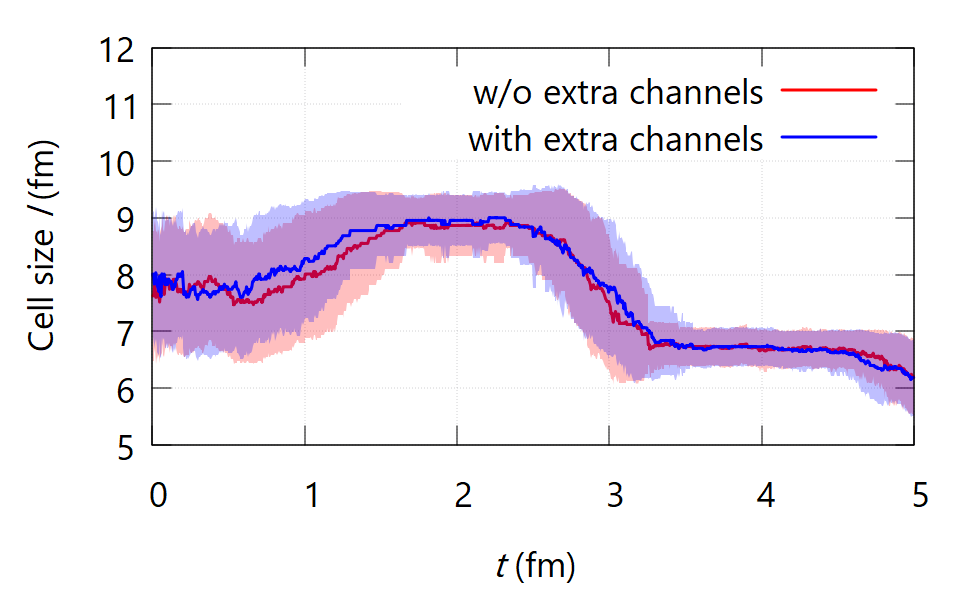}
  \caption{Cell size for maximum entropy across the time evolution. Without extra channels case is shown in red, and with extra channels case is shown in blue, both evolve similarly with the maximum plateau formed around $t$ = 2 fm.}
  \label{fig:expanding_cell_size_entropy}
\end{figure}

The result in Fig~\ref{fig:expanding_cell_size_entropy} shows a clear tendency of the cell size against time, which is insensitive to the interaction channels. This tendency suggests that the medium goes through 3 distinct phases during its time evolution.

\begin{enumerate}
  \item At small $t$, we see that a smaller cell size maximizes Shannon entropy, this can be interpreted as the algorithm capturing the microscopic features of the multiple small hotspots created at the beginning.
  \item At around $t$ = 2 fm, we see a plateau after an increase. This is reached when the energy of each small hotspot has spread and amalgamated into a single large cluster with relatively more homogeneous energy density compared to the early time.
  \item At $t$ larger than 3 fm. The optimal cell size decreases and forms another plateau. This might be caused by the non-uniform cooling down of the medium, thus causing the formation of multiple smaller clusters of relatively high energy density inside the main cluster.
\end{enumerate}

Regardless of the different behavior compared to the monotonic increase from the previous algorithm, we still see consistency in terms of the similarity between the with and without extra channels cases and the general size of the characteristic length of the medium compared to the previous method.

\subsubsection{Knudsen Number Result}

\begin{figure}[h]
  \includegraphics[width=\linewidth]{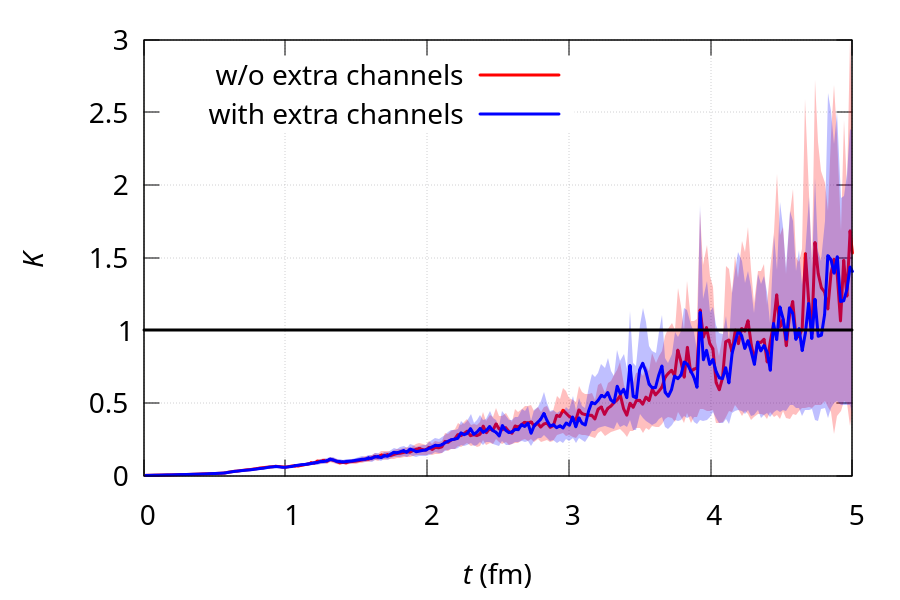}
  \caption{Knudsen number up to 5 fm compared to the hydrodynamic limit $K_{n} = 1$ (black solid line) calculated using Shannon Entropy maximization characteristic length. Both with extra channels (blue) and without extra channels (red) cases increase rapidly above the hydrodynamic limit at the later time, thus indicating a breakdown of hydrodynamization which is consistent with a recent kinetic theory model considering only elastic collisions \cite{Nugara:2024net, Nugara:2025ueb}.}
  \label{fig:expanding_knudsen}
\end{figure}

We see from Fig~\ref{fig:expanding_knudsen} that the Knudsen number starts well below 1 at the early time but then deviates from the hydrodynamic limit at the later time. Since from Fig~\ref{fig:mean_radius}, we see only a small change in the characteristic medium length, hence it implies that this increase is due to the rapidly increasing mean free path length at the later time. Although we refer to the Shannon Entropy maximization characteristic length to calculate Knudsen number, the result with the expectation maximization method characteristic length should give a similar value as it yields the characteristic length between 8-11 fm.

This breaking away from the hydrodynamic regime is also observed in the other work with effective kinetic theory in a 3D expanding system with only elastic collisions \cite{Nugara:2024net, Nugara:2025ueb}. Understandably, the Knudsen number went above 1 at a much later time when it is already well past the typical hydrodynamization time of around 1 fm. Therefore, this indicates the limitation of the kinetic theory at later time. On the other hand, we see Knudsen number exponentially increases much faster than what was observed in the case without inelastic collisions \cite{Nugara:2025ueb}, referring back to Fig.~\ref{fig:expanding_cell_size_entropy}, the characteristic length scale only varies between 6-9 fm, thus the rapid increase in Knudsen number is attributed more to the increase in mean free path. As inclusion of inelastic collisions allowed the number of particles to increase with time, the average energy carried per particle is much lower than compared to the elastic-collision only case. The lower average energy per particle then caused the average cross-section to quickly diminish with time, hence the exponential growth in the $\lambda_{\text{mfp}}$. Nevertheless, inelastic collision is an important driving factor for thermal and chemical equilibration, while its addition also caused the Knudsen number to rise faster than the elastic-collision-only case. Therefore, it is clear that how we calculate the inelastic cross-section plays an important role between the speed of equilibration at the early time and the breaking away from hydrodynamic limit at the later time.

\section{Conclusion}\label{sec:conclusion}

In this study, we have analyzed the thermalization, chemical equilibration, and hydrodynamization behavior in Au-Au at 200 GeV collision energy with a parton cascade model including gluon production and absorption from quark interaction. We found in the box simulation that the extra channels accelerate thermalization and chemical equilibration, hence the effect from these channels is not negligible. In contrast to the box simulation, we did not see a meaningful difference in the expanding medium case. Moreover, our result suggests equilibration time around 0.2 fm and momentum isotropization time around 2 fm, which are on the same order of the common hydrodynamic model initialization time of 0.2 - 1 fm. Lastly, we found that the Knudsen number rapidly increases above 1 after 4 fm which suggests a breakaway from the hydrodynamic regime at the later time.

We first expanded SMASH hadron transport model to the partonic case. We included 2-to-2 and 2-to-3 interactions between quarks and gluons. The 2-to-3 cross sections are calculated using the improved Gunion-Bertsch approximation up to the tree-level diagram. We fixed the coupling constant while using the dynamic Debye Mass to avoid infrared divergence. We have proved that the model indeed satisfies the detailed balance principle and maintains a stable state when the medium starts already in the equilibrium state defined by the Boltzmann distribution function.

One of the main distinctions of our model is the addition of gluon absorption and radiation when interacting with quarks. Previous models often only consider gluon absorption and radiation either through decay or 2-to-3 interactions exclusively to gluons because the initial state of the system is gluon-dominated. However, as we will see in the following result, the transition from gluon-dominated into quark-dominated system happens much sooner than the equilibration time scale. Thus, the gluon absorption and radiation through the interaction with quarks might play an important role in the equilibration.

In order to ensure that the simulation behaves as expected, we simulate the system with and without the extra channels qg → qgg and qq → qqg with 50 events for each case to see the effect of gluon absorption and radiation from quark interactions. As we have confirmed that the medium is stable once it reaches the equilibrium state. We proceeded to examine whether the medium shall converge toward the equilibrium state when initialized far from the equilibrium. We initialized far-from-equilibrium state by sampling the initial spatial distribution from the Glauber model and the initial momentum distribution from the deep inelastic scattering with transverse momentum larger than 2 GeV.

The box equilibration simulation showed that indeed starting from the aforementioned far-from-equilibrium state, the medium then evolves toward the equilibrium state. Moreover, we also see the addition of qg → qgg and qq → qqg collision channels speeds up the equilibration rate. Momentum isotropy is reached first at 2 fm followed by thermalization at around 3 fm, and lastly, chemical equilibrium is reached far later around 30 fm. Even though the extra channels did not show much difference in hydrodynamization, a clear difference is seen in number density as the case with the extra channels already saturated at the equilibrium value around 30 fm while without extra channels case took even more than 50 fm to reach the same number density. Measured temperature also shows similar behavior except that the extra channels case speeds up the equilibration time. The main point of this simulation is to show that the medium indeed converges toward the equilibrium state even when initialized far from the equilibrium.

After we have defined how to identify equilibration in the box simulation case, we then apply the same equilibration criteria in the expanding medium case. Although we see that momentum distribution converges toward the Boltzmann distribution almost immediately, we see that it then deviates at the later time. Chemical equilibration also cannot be clearly defined due to the fugacity peaking at the early time before rapidly decreasing afterwards. This is in part due to the difficulty in defining the volume of the medium. In contrast to the box simulation, here we do not see a clear effect from the addition of the extra channels. This is due to the parton number where we have a small number of quarks, while the extra channels are to describe the gluon absorption and radiation from quark interactions.

On the other hand, we clearly see that temperature only stabilize after around 1 fm while momentum isotropy is reached around 2 fm time mark. Although we cannot see a clear equilibration behavior, the condition for the switch to the hydrodynamic model is already satisfied around this timescale. The deviation of the momentum distribution from the Boltzmann distribution function and the rapid decrease of the momentum after 3 fm implies that kinetic theory no longer simulates the medium well at this stage, hence the need to switch to a hydrodynamic description. 

It is commonly assumed that the equilibrium state cannot be reached when the system does not have enough particle density, consequently, a better distinction in the equilibration stage might be clearer to see if we look at a medium with higher multiplicity from a larger collision system or a higher collision energy. On the other hand, we may also go to the low multiplicity case such as in the small system where, in contrast to the high multiplicity case, the indicators for equilibration and hydrodynamization might become obscured. This is a good test to see whether we can still see the same indicators of equilibration and hydrodynamization that we have established in the higher multiplicity case, this also serves as a test of the limit of the kinetic theory approach. A different approach can also be taken to expand this study by using this model as a base to construct an initial state for a hydrodynamic model. Although, in this case, we will first need to reconcile the mismatch between the hydrodynamization time scale ($\sim 10^{0}$ fm) compared to the common initial time used in hydrodynamic models ($\sim 10^{-1}$ fm) which we reckon as a long-standing problem in the kinetic theory approach.

\section{Acknowledgement}

We would like to thank Steffen A. Bass from Duke University, Hannah Elfner and Jan Staudenmaier from Frankfurt University for the invaluable discussions and counsels throughout the course of the research.
The numerical computation in this work was carried out at the Yukawa Institute Computer Facility.
This work was also supported by JSPS KAKENHI Grant Numbers, JP20H00156, JP20H11581, JP2500449 (C.N.) and by the World Premier International Research Center Initiative (WPI) under MEXT, Japan (C.N.).

\appendix
\section{$\text{R}^{2}$-value and homoskedasticity}\label{appendix:r2_homosked_explanation}

The coefficient of determination ($\text{R}^{2}$) and homoskedasticity are both important measures of how well a mathematical model describes a dataset based on the explained variances. $\text{R}^{2}$-value is given as

\begin{equation}
    \text{R}^{2} = 1 - \frac{\sum_i{(y_{i}^{data} - y_{i}^{model})^{2}}}{\sum_{i}{(y_{i}^{data} - y_{mean})^{2}}}
\end{equation}
expresses how much the model explains the variation of the observed data. $\text{R}^{2} = 1$ means that the model explains the variation in the data and fits the data perfectly. $\text{R}^{2} = 0$ means that the model performs as well as simply taking the average value of the data, while a negative $\text{R}^{2}$ means that the model performs even worse to describe the data than just a single constant value, which is the average value.

On the other hand, homoskedasticity is a measure of how constant the residuals are around zero. Homoskedasticity signifies that after subtracting the model value from the data, we will only be left with Gaussian white noise. Hence, any analysis on the fitness of the model will only hold any values when homoskedasticity is present, high $\text{R}^{2}$ value with heteroscedastic residual means that we might have missed an important dependent variable in the model, thus, the error measured based on that model is untrustworthy and the resulting $\text{R}^{2}$ value is misleading. In this paper, we employed the Breusch-Pagan test to measure homoskedasticity.

As it is often understood that only the lower energy level region in the energy spectrum follows the Boltzmann distribution function, we first have to define the upper energy limit to which we apply the linear regression fit. In order to define this upper energy limit, we apply regression analysis and scan for different upper energy limits after 50 fm time evolution from a box simulation to give enough time for the energy spectrum to converge toward the Boltzmann distribution and settle on a certain value for its slope.

\begin{figure}[H]
  \includegraphics[width=\linewidth]{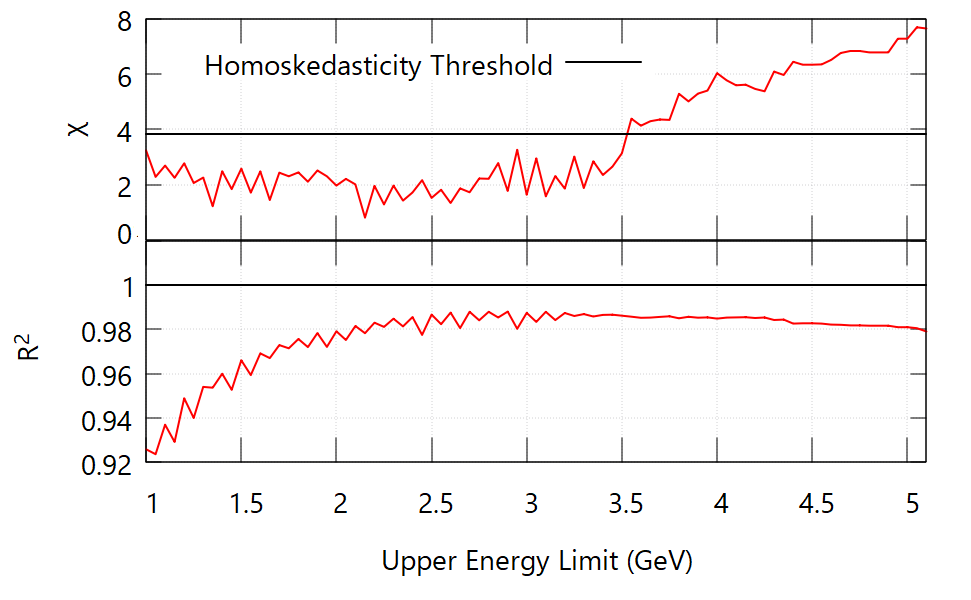}
  \caption{Regression analysis from the upper energy limit scan. The quality of the Boltzmann distribution fit to the energy spectrum quantified in $\chi$ (top) and $\text{R}^{2}$-value (bottom) with respect to the upper energy limit used in the data fit. Rapid increases in $\chi$ after 3 GeV show that the Boltzmann distribution cannot fit the energy spectrum well after 3 GeV.}
  \label{fig:box_equilibration_e_lim}
\end{figure}

Figure \ref{fig:box_equilibration_e_lim} shows $\text{R}^{2} \text{-value}$ forms a wide curve with the peak around 3 GeV and slowly decreases afterwards. Similarly, the measured $\chi < \text{3.841}$ also shows a sudden rapid increase around the same value. Hence, we assume that the best region to apply linear regression fit is between $E$ = 0 GeV until $E$ = 3 GeV to find the temperature of the medium.

\bibliography{Bibliography.bib}

\end{document}